\documentclass[%
 reprint,
 amsmath,amssymb,amsthm,
 nofootinbib,
 aps,
]{revtex4-1}

\usepackage{dcolumn}
\usepackage{bm}
\usepackage[mathlines]{lineno}
\usepackage{graphicx}
\graphicspath{ {./images/} }
\usepackage{csquotes}
\usepackage[final]{changes}

\begin{document}
\preprint{APS/123-QED}

\title{Quantum Theory of Electron Spin Based on the Extended Least Action Principle and Information Metrics of Spin Orientations}

\author{Jianhao M. Yang}
\email[]{jianhao.yang@alumni.utoronto.ca}
\affiliation{Qualcomm, San Diego, CA 92321, USA}

\date{\today}		

\begin{abstract}
Quantum theory of electron spin is developed here based on the extended least action principle and assumptions of intrinsic angular momentum of an electron with random orientations. The novelty of the formulation is the introduction of relative entropy for the random orientations of intrinsic angular momentum when extremizing the total actions. Applying recursively this extended least action principle, we show that the quantization of electron spin is a mathematical consequence when the order of relative entropy approaches a limit. In addition, the formulation of the measurement probability when a second Stern-Gerlach apparatus is rotated relative to the first Stern-Gerlach apparatus, and the Schr\"{o}dinger-Pauli equation, are recovered successfully. Furthermore, the principle allows us to provide an intuitive physical model and formulation to explain the entanglement phenomenon between two electron spins. In this model, spin entanglement is the consequence of the correlation between the random orientations of the intrinsic angular momenta of the two electrons. Since spin orientation is an intrinsic local property of the electron, the correlation of spin orientations can be preserved and manifested even when the two electrons are remotely separated. The entanglement of a spin singlet state is represented by two joint probability density functions that reflect the orientation correlation. Using these joint probability density functions, we prove that the Bell-CHSH inequality is violated in a Bell test. To test the validity of the spin-entanglement model, we propose a Bell test experiment with time delay. Such an experiment starts with a typical Bell test that confirms the violation of the Bell-CHSH inequality but adds an extra step that Bob's measurement is delayed with a period of time after Alice's measurement. The present theory suggests that the Bell-CHSH inequality becomes non-violated if the time delay is sufficiently large. 
\end{abstract}

\maketitle
\section{Introduction}
In quantum mechanics, the spin of an electron is one of the most important physical observables to demonstrate the essence of quantum theory. For instance, measurement of the spin of an electron always obtains two possible values, spin-up or spin-down, along the direction of the measuring magnetic field; To describe the dynamics of an electron with spin, multi-components wave functions in the Schr\"{o}dinger-Pauli equation are needed; In investigating the quantum entanglement phenomenon, the singlet state of an electron pair is often used as example to illustrate conceptual challenges such as in the Bohm version EPR thought experiment~\cite{EPR, Bohm1957}, or to verify the violation of Bell inequality~\cite{Bell}. However, these important quantum features are either not derived from first principles or still have conceptual difficulties. Quantization of spin measurement outcome is normally introduced as a postulate in standard quantum mechanics. Whether the wave function is epistemological or associated with physical reality is still under debate. The violation of Bell inequality confirms that spin correlation can be non-local in the sense that the correlation is inseparable even if the two electrons are space-like separated. Without an intuitive physical model, this inseparability is not possible to comprehend in classical terms and still puzzles the physics community~\cite{Brunner}. A more intuitive physical model and a clearer theory of electron spin to better explain these challenges are still much desirable in modern quantum physics.

Historically, a number of physical models for electron spin have been proposed~\cite{Tomonaga, Sebens}. Here we will briefly review two prominent ones. The first model considers the electron as a rotating rigid body with a uniform distributed charge, and has been adopted in a stochastic mechanism~\cite{Dankel1977, Fairs1982, Beyer}. This model can derive the relation between the angular momentum and the magnetic moment of an electron. By applying the theory of stochastic mechanics, the electron spin is treated as the averaged of hidden angular momentum variables with random orientations. Formulations equivalent to quantum mechanics can be recovered. However, one of the problems of this model is that when the radius of the rigid body is small enough, the edge moves faster than the speed of light~\cite{Tomonaga}. In addition, the velocity fields in stochastic mechanics retain the non-local characteristics, a drawback that should be avoided. Another model is the so-called ``Zitterbewegung" model where the origin of spin is proposed as a result of the helical motion of a "light-like" particle~\cite{Wey1947, Barut1984, Niehaus16, Niehaus22}. The particle is moving automatically in a circle with the speed of light. There is also a simpler model that considers an electron as a point particle with intrinsic magnetic moment and intrinsic angular momentum. Such a model does not offer an explanation as to how the intrinsic angular moment or the intrinsic magnetic moment originates. 

Recent interest in searching for foundational principles of quantum theory from the information perspective~\cite{Rovelli:1995fv, zeilinger1999foundational, Brukner:ys, Brukner:1999qf, Fuchs2002, brukner2009information, spekkens2007evidence, Spekkens:2014fk, Paterek:2010fk, gornitz2003introduction, lyre1995quantum, Hardy:2001jk, masanes2011derivation, Mueller:2012ai, Masanes:2012uq, chiribella2011informational, Mueller:2012pc, Hardy:2013fk, kochen2013reconstruction, 2008arXiv0805.2770G, Hall2013, Hoehn:2014uua, Hoehn:2015,  Caticha2011, Caticha2019, Frieden, Reginatto} can offer new perspectives for the investigation of spin theory. The motivation here is to reformulate quantum mechanics with information-theoretic principles such that some of the conceptual challenges can be resolved. Inspired by this program, an extended least action principle has been proposed~\cite{Yang2023}. The principle extends the classical least action principle by incorporating information metrics on vacuum fluctuations. Based on this principle, non-relativistic quantum mechanics~\cite{Yang2023} and quantum scalar field theory~\cite{Yang2024} have been recovered, demonstrating the general applicability of the theoretical framework. 

The goal of this paper is to apply the extended least action principle to develop the theory for electron spin with the goal of providing new insights into some of the challenges mentioned earlier. This is achieved by introducing an additional assumption of intrinsic angular momentum with random orientations and choosing appropriate relative entropy for the random orientations. We will show that the results are indeed quite fruitful. Discreteness of measurement result of electron spin is a mathematical result when the order of relative entropy approaches a limit. The formulation also gives equivalent results as standard quantum mechanics when the direction of magnetic field of a second Stern-Gerlach apparatus is rotated with an angle from the direction of magnetic field of the first Stern-Gerlach apparatus. Recursively applying the extended least action principle, we can derive the Schr\"{o}dinger-Pauli equation.

With respect to spin entanglement between two electron spins, we will demonstrate that the origin of entanglement is attributed to the correlation between the random variables, the orientations of the intrinsic angular momenta, of the two electrons. The correlation can be established during preparation. The correlation can be preserved even though the two electrons are remotely separated and without interaction since the orientation of the angular momentum is an intrinsic local property of the electron. Mathematically, we give an equivalent formulation of the four Bell states using the probability density function of the angular momentum orientation, instead of the wave function in standard quantum mechanics. Based on the probability density function of a spin singlet state, the Bell-CHSH inequality is proved to be violated due to the same root cause of entanglement - the correlation between the random orientations of the intrinsic angular momenta. Specifically, since this correlation is preserved even when the two electrons are separated, the joint probability for the measurement outcome cannot be factorized into a product of two individual terms, a requirement for the Bell inequality to hold.

Two potential experiments are proposed to test the validity of the theory and its difference from standard quantum mechanics. In the first experiment, the Stern-Gerlach apparatus is configured with a sufficiently weak gradient of the inhomogeneous magnetic field along the $z$-axis; We expect the electron detector screen to exhibit a continuous distribution of displacements along the $z$-axis around two areas, instead of only two discrete lines. The second experiment is to modify a typical Bell test experiment that has already confirmed the violation of the Bell-CSHS inequality. Here, we purposely let Alice and Bob perform their measurements at different times on each entangled pair of electrons. While Alice performs her measurement at time $t_a$, Bob intentionally delays his measurement at time $t_b=t_a+\Delta t$. The present theory suggests that when $\Delta t$ is chosen properly, the Bell-CSHS inequality will not be violated.

The paper is organized as follows. In Section \ref{LIP}, we review the extended least action principle and its underlying assumptions. Detailed calculations showing how the basic quantum theory is derived from the principle are provided in Appendix \ref{sec:shorttime}. Section \ref{sec:SpinTheory} is devoted to developing the spin model, where the extended least action principle is incorporated with the relative entropy for the spin random orientations, resulting in a probability density function for the spin orientation. The probability density function becomes two Dirac delta functions when the order of relative entropy approaches a limit, corresponding to the quantization of measurement outcomes. The theory for rotation of the spin is developed in this section as well. In Section \ref{sec:PauliEq}, the Schr\"{o}dinger-Pauli equation is derived by combining the effects of random translational fluctuations and random rotational fluctuations when applying the extended least action principle. Section \ref{sec:entanglement} develops the theory for spin entanglement based on the correlations between the random orientations of intrinsic angular momenta, which allows us to prove that the Bell inequality is violated for a singlet state. To test the validity of the theory, two potential experiments are proposed in Section \ref{sec:PossibleExp}. 

There are clear limitations in the present theory, as discussed in Section \ref{sec:discussion}, particularly the assumption that the order of relative entropy increases when the electron travels along an inhomogeneous magnetic field. We give some intuitive explanations in Section \ref{sec:discussion}, but a more rigorous theory must be developed to further justify these assumptions.

\section{The Extended Least Action Principle}
\label{LIP}
The theoretical framework in this paper is developed based on the extended least action principle proposed in~\cite{Yang2023}. We will briefly review the principle before introducing additional assumptions in order to develop the spin theory. In \cite{Yang2023}, the least action principle in classical mechanics is extended to derive the quantum formulation by factoring in the following two assumptions.
\begin{displayquote}
\emph{Assumption 1 -- A quantum system experiences vacuum fluctuations constantly. The fluctuations are local and completely random.}
\end{displayquote}
\begin{displayquote}
\emph{Assumption 2 -- There is a lower limit to the amount of action that a physical system needs to exhibit in order to be observable. This basic discrete unit of action effort is given by $\hbar/2$ where $\hbar$ is the Planck constant.}
\end{displayquote}

The first assumption is generally accepted in mainstream quantum mechanics and is responsible for the intrinsic randomness of the dynamics of a quantum object. Although we do not know the physical details of the vacuum fluctuation, the crucial assumption here is the locality of the vacuum fluctuation. This implies that for a composite system, the fluctuation of each subsystem is independent of each other.  

The implications of the second assumption need more elaborations. Historically, the Planck constant was first introduced to show that the energy of radiation from a black body is discrete. One can consider the discrete energy unit as the smallest unit to be distinguished, or detected, in the black-body radiation phenomenon. In general, it is understood that Planck constant is associated with the discreteness of certain observables in quantum mechanics. Here, we instead interpret the Planck constant from an information measure point of view. Assumption 2 states that there is a lower limit to the amount of action that the physical system needs to exhibit in order to be observable or distinguishable in potential observation, and such a unit of action is determined by the Planck constant. 

Making use of this understanding of the Planck constant conversely provides us a new way to calculate the additional action due to vacuum fluctuations. That is, even though we do not know the physical details of vacuum fluctuations, the vacuum fluctuations manifest themselves via a discrete action unit determined by the Planck constant as an observable information unit. If we can define an information metric that quantifies the amount of observable information manifested by vacuum fluctuations, we can then multiply the metric by the Planck constant to obtain the action associated with vacuum fluctuations. Then, the challenge of calculating the additional action due to the vacuum fluctuation is converted to define a proper new information metric $I_f$, which measures the additional distinguishable, and hence observable, information exhibited due to vacuum fluctuations. The problem of defining an appropriate information metrics becomes less challenging since there are information-theoretic tools available. 

The first step is to assign a transition probability distribution due to vacuum fluctuation for an infinitesimal time step at each position along the classical trajectory. The distinguishability of vacuum fluctuation can then be defined as the information distance between the transition probability distribution and a uniform probability distribution. The uniform probability distribution is chosen here as a reference to reflect the complete randomness of vacuum fluctuations. In information theory, the common information metric for measuring the distance between two probability distributions is relative entropy. Relative entropy is more fundamental than Shannon entropy since the latter is just a special case of relative entropy when the reference probability distribution is a uniform distribution. However, there is a more important reason to use relative entropy. As shown in later sections, when we consider the dynamics of the system for an accumulated time period, we assume that the initial position is unknown but given by a probability distribution. This probability distribution can be defined along the position of a classical trajectory without vacuum fluctuations or with vacuum fluctuations. The information distance between the two probability distributions gives additional distinguishability because of vacuum fluctuations. It is again measured by a relative entropy. Thus, relative entropy is a powerful tool allowing us to extract meaningful information about the dynamic effects of vacuum fluctuations. The concrete form of $I_f$ will be defined later as a functional of relative entropy that measures the information distances of different probability distributions caused by vacuum fluctuations. Thus, the total action from classical path and vacuum fluctuation is
\begin{equation}
\label{totalAction}
    S_t = S_c + \frac{\hbar}{2}I_f,
\end{equation}
where $S_c$ is the classical action. Non-relativistic quantum theory can be derived through a variation approach to minimize such a functional quantity, $\delta S_t=0$. When $\hbar \to 0$, $S_t=S_c$. Minimizing $S_t$ is then equivalent to minimizing $S_c$, resulting in Newton's laws in classical mechanics. However, in quantum mechanics, $\hbar \ne 0$, the contribution of $I_f$ must be included when minimizing the total action. We can see that $I_f$ is where the quantum behavior of a system comes from. These ideas can be condensed as follows.
\begin{displayquote}
\emph{\textbf{Extended Least Action Principle} -- The law of physical dynamics for a quantum system tends to exhibit as little as possible the action functional defined in (\ref{totalAction}).}
\end{displayquote}

Non-relativistic quantum formulation can be derived based on the extended least action principle if we only consider the translational component of vacuum fluctuations~\cite{Yang2023}. A brief description of the derivation is given in Appendix \ref{sec:shorttime} for self-reference. It has also been shown that the quantum scalar field theory can be formulated based on the principle~\cite{Yang2024}, which further demonstrates its general applicability. In this paper, we will apply the extended least action principle to derive the quantum theory of electron spin by considering not only translational, but also rotational components of vacuum fluctuations. 

\section{Quantum Theory of Spin} 
\label{sec:SpinTheory}

\subsection{Spin Model}
\label{sec:entBipartite}
In order to explain the existence of electron spin, additional assumptions on electron properties are needed. Historically, several models have been proposed to explain the existence of spin. The most popular model is to consider the electron as a rotating rigid body with uniform distributed charge. Denote the magnetic moment of the electron as $\Vec{\mu}$ and the angular momentum as $\Vec{L}$. The relation between $\Vec{\mu}$ and $\Vec{L}$ can be derived if we consider the electron as a localized electrical distribution with current density $\Vec{j}(\Vec{r})$, and the magnetic moment of electron is the magnetic dipole moment, which is calculated as
\begin{equation}
    \Vec{\mu} = \frac{1}{2} \int \Vec{r}\times \Vec{j}(\Vec{r}) d^3 \Vec{r}.
\end{equation}
Assuming the electrical density and mass density is identical, the above expression can be rewritten as 
\begin{equation}
\label{muL}
    \Vec{\mu} = -\frac{e}{2m} \int \Vec{r}\times \Vec{p}(\Vec{r}) d^3 \Vec{r} = -\frac{e}{2m}\Vec{L},
\end{equation}
where $\Vec{p}(\Vec{r})$ is the momentum density. However, one of the problems of this model is that when the radius of the rigid body is small enough, the edge moves faster than the speed of light~\cite{Tomonaga}. There is also a simpler model that considers an electron as a point particle with intrinsic magnetic moment and intrinsic angular momentum, and they still satisfy the relation (\ref{muL}). Such a model does not offer an explanation as to how the intrinsic angular moment or the intrinsic magnetic moment originates. 

Based on the success of applying the extended least action principle in deriving quantum theories~\cite{Yang2023, Yang2024}, we wish to explain the origin of the intrinsic magnetic moment and angular momentum by applying the same principle but adding additional refinement on the assumption of the vacuum fluctuations in Section \ref{LIP}. The electron exhibits random displacement as a result of vacuum fluctuation, and this displacement is described as a vector. The displacement vector is intrinsically local in the sense that it is independent of the reference of origin of the coordinate system. If we further assume that the displacement vector contains not only translational components, but also rotational components, then the electron will exhibit intrinsic angular momentum as a result of such random rotation. In Appendix \ref{appendix:intrinsicSpin}, using the extended least action principle, we show that if the rotation component of the displacement vector is a circular movement, the averaged magnitude of intrinsic angular momentum turns out to be $ L_s  = \hbar/2$, while the orientations of intrinsic angular momentum are completely random. That is, the intrinsic angular momentum is isotropic and its orientation randomly fluctuates. Although our model is very different from the rigid body model, we assume that the relation \eqref{muL} still holds except with a factor $g_s$ to be determined further
\begin{equation}
    \label{muLs}
    \Vec{\mu} = - g_s\frac{e}{2m}\Vec{L}_s.
\end{equation}
Since the orientation of the magnetic moment is always opposite to the orientation of the intrinsic angular momentum, the orientation of the intrinsic magnetic moment fluctuates randomly as well. Although the result in Appendix \ref{appendix:intrinsicSpin} is impressive enough, the assumption of circular motion of displacement vectors due to vacuum fluctuation is rather strong\footnote{It is desirable to develop a better model than that is described in Appendix \ref{appendix:intrinsicSpin} to derive the $\hbar/2$ magnitude of intrinsic angular momentum, but we leave it as a future research topic.}. Instead, we will just assume the existence of an intrinsic angular momentum of magnitude $\hbar/2$ and with a random orientation for an electron. In summary, in addition to Assumptions 1 and 2 as described in Section \ref{LIP}, we need
\begin{displayquote}
\emph{Assumption 3 -- An electron has intrinsic angular momentum with magnitude of $\hbar/2$ and completely random orientation in free space.}
\end{displayquote}
This additional assumption allows us to apply the extended least action principle to derive the probability density of the orientation of the intrinsic angular momentum. 

Suppose that we choose a reference frame such that the electron is at rest, that is, the average translational momentum is zero. The electron still exhibits an intrinsic angular momentum with random orientation according to Assumption 3. We want to derive the probability distribution of the intrinsic angular momentum orientations. If no external magnetic field is applied, the probability distribution is simply a uniform distribution according to Assumption 3. Now, suppose that an external magnetic field along the direction of the $z$-axis, $B_z$, is applied, the probability distribution of the magnetic moment orientations is no longer uniform. We show next that this probability distribution can be derived from the extended least action principle. 

Due to the interaction of the electron magnetic moment and the external magnetic field, the electron is experiencing Larmor precession around the $z$-axis, as shown in Figure 1. Denote the probability density of the intrinsic angular momentum orientations as $p(\theta, \varphi)$ where $\theta$ is the angle between the direction of $\Vec{L}_s$ and the $z$-axis, and $\varphi$ is the angle between the projection of $\Vec{L}_s$ in the X-Y plane and the $x$-axis. There is no reason to assume that the probability density $p$ depends on $\varphi$ since the external field is along the $z$-axis. We can simply drop the parameter $\varphi$ from $p(\theta, \varphi)$. The Larmor angular frequency is $\omega=eB_z/m$, which is independent of angle $\theta$. The corresponding potential energy for Larmor precession is
\begin{equation}
\begin{split}
    U &= -\Vec{\mu}\cdot\Vec{B} \\
    &=\frac{e}{2m}g_sB_z L_s \cos(\theta)\\
    &=\frac{1}{2}g_s\omega L_s\cos(\theta).
\end{split}
\end{equation}
where $L_s=\hbar/2$ is the magnitude of the intrinsic angular momentum. The above expression allows us to calculate the expectation value of classical action for an infinitesimal time period $\Delta t$ as
\begin{equation}
\label{spinAction}
    \begin{split}
        A_c &= -\int_0^{\pi}\int_0^{\Delta t} p(\theta)U d\theta dt \\
        & = -\frac{1}{2}g_s\int_0^{\pi}\int_0^{\Delta t} p(\theta)\omega L_s\cos(\theta)d\theta dt.
    \end{split}
\end{equation}
Note that $d\varphi = \omega dt$ where $\varphi$ is the angle of circular rotation due to Larmor precession. In an infinitesimal period of time $\Delta t$, the intrinsic angular momentum of the electron rotates $\Delta\varphi = \omega\Delta t$. We can rewrite
\begin{equation}
\label{spinAction2}
    A_c = -\frac{1}{2}g_sL_s\int_0^{\pi}\int_0^{\Delta \varphi} p(\theta) \cos(\theta)d\theta d\varphi.
\end{equation}
\begin{figure}
\begin{center}
\includegraphics[scale=0.75]{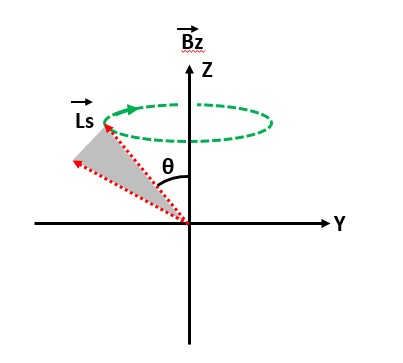}
\caption{Precession of the intrinsic angular momentum $\Vec{L}_s$ around the magnetic field $\vec{B}_z$. The shading area indicates that the orientation of $\Vec{L}_s$ is randomly fluctuating.}
\end{center}
\label{fig:precession}
\end{figure}

To compute $I_f$, we recall that the framework based on the extended least action principle~\cite{Yang2023} allows us to choose a general relative entropy definition such as Kullback-Leibler divergence, R\'{e}nyi divergence, or Tsallis divergence~\cite{Tsallis, Erven2014, Nielsen2011}. \added {The R\'{e}nyi divergence, or the Tsallis divergence, is the one-parameter generation of the K-L divergence, where the parameter is called the order of divergence. The K-L divergence is a special limit when the order of divergence approaches one. As will be seen later, the parameter of the order of divergence is needed to explain spin quantization. Therefore, we exclude the K-L divergence and can choose either the R\'{e}nyi divergence or the Tsallis divergence. There is no fundamental reason to choose one or the other, although the Tsallis divergence has the advantage of mathematical simplicity. In Appendix \ref{sec:choosingI_f}, we prove that choosing R\'{e}nyi divergence gives the similar results on spin quantization.} With these considerations, we define $I_f$ for the infinitesimal period of time $\Delta t$ as follows. 
\begin{equation}
\label{I_f}
    I_f = \frac{1}{\alpha-1}\{\int_0^{\pi}\int_0^{\Delta \varphi} \frac{p^{\alpha}(\theta)}{\sigma^{\alpha-1}}d\theta d\varphi - 1 \},
\end{equation}
where $\alpha \in (0, 1) \cup (1, \infty)$ is the order of Tsallis divergence, and $\sigma$ is an uniform probability density to reflect the total ignorance of knowledge due to complete randomness of orientations. Then, the total action as defined in (\ref{totalAction}) is
\begin{equation}
\label{totalActionT}
\begin{split}
    A_t &= -\frac{1}{2}g_sL_s\int p(\theta) \cos(\theta)d\theta d\varphi \\
    & + \frac{\hbar}{2(\alpha-1)}\{\int\frac{p^{\alpha}(\theta)}{\sigma^{\alpha-1}}d\theta d\varphi - 1 \}.
\end{split}
\end{equation}
Taking the variation of $A_t$ over the functional variable $p(\theta)$, and demanding $\delta A_t = 0$, we obtain
\begin{equation}
    -\frac{1}{2}g_sL_s\cos(\theta) + \frac{\alpha\hbar}{2(\alpha-1)}[\frac{p(\theta)}{\sigma}]^{\alpha -1} =0.
\end{equation}
This gives
\begin{equation}
    \label{ptheta}
    \begin{split}  
    p(\theta) &= \sigma [\frac{(\alpha-1)g_sL_s}{\alpha\hbar}\cos(\theta)]^{\frac{1}{\alpha -1}} \\
    &=\frac{1}{Z_{\alpha}}[\cos(\theta)]^{\frac{1}{\alpha -1}},
    \end{split}
\end{equation}
where $Z_{\alpha}$ is a normalization factor that is dependent on $\alpha$. Now, for $p(\theta)$ to be a probability density number, it must be real and non-negative. This imposes restrictions on the value of $\alpha$. Since $\cos(\theta)$ can be a negative number, $1/(\alpha - 1)$ should be an even integer. Let $1/(\alpha - 1) = 2m$ where $m \in \mathbb{N}$. This gives
\begin{equation}
\label{alpham}
    \alpha = 1 + \frac{1}{2m}.
\end{equation}
Thus, the probability density is rewritten as
\begin{equation}
    \label{pthetam}
    p_m(\theta)=\frac{1}{Z_m}[\cos(\theta)]^{2m}, m \in \mathbb{N}.
\end{equation}
Thus, from the extended least action principle and Assumption 3, we obtain a family of probability density functions $p_m(\theta)$. Mathematically, each of the probability density functions is a legitimate solution. We will further impose physical constraints in the next subsection and single out the solution that can explain the measurement results of the electron spin.

\subsection{Discreteness of Spin Measurement Results}
\label{sec:entBipartite} 
The probability density functions $p_m(\theta)$ in (\ref{pthetam}) are valid only relative to the context of the magnetic field $B_z$ since $B_z$ defines the direction of the space to measure the spin. Without the external magnetic field, i.e., $B_z = 0$, then the spin orientation is completely random. This corresponds to the case $m=0$ (or $\alpha\to\infty)$ such that $p_m(\theta)$ is a uniform distribution. Recognizing the fact that spin measurement result is context dependent is itself an important result\footnote{More precisely, we should label the probability density functions as $p_m[(\theta) | B_z]$ to show its contextual dependency. For simplicity of notation, we will not adopt such labeling in this paper. However, the conceptual implications on this contextual dependency will be discussed in Section \ref{sec:discussion}.}.

When an inhomogeneous field $B_z$ is applied in the Stern-Gerlach experiment, the electron movement is no longer randomly distributed along the $z$ -axis. This means that we need to choose $p_m(\theta)$ with $m$ larger than zero. However, there are infinite numbers of $m > 0$ to choose from. We only know that the value of $m$ depends on the inhomogeneous field $B_z$. The exact relation between the value of $m$ and the inhomogeneous field $B_z$ cannot be derived from the extended least action principle. At this point, we need to make another assumption on the relation between the value of $m$ and the inhomogeneous field $B_z$, that $m$ monotonically increases as the electron travels along the inhomogeneous field $B_z$. The justification of this assumption requires a more detailed physical model of electron. However, we will provide an intuitive explanation of this assumption and a possible physical model in Section \ref{subsec:limitations}. The intention here is to make minimal assumptions so that we can recover the quantum theory of the electron spin, including the discreteness of spin measurement results, the Pauli-Schrodinger equation, and entanglement between two correlated electron spins, as shown in later sections. Explicitly, we have
\begin{displayquote}
\emph{Assumption 4 -- Parameter $m$ in (\ref{pthetam}) increases monotonically as the electron travels along an inhomogeneous magnetic field.}
\end{displayquote}
With this assumption, one can extrapolate the outcomes if the gradient of $B_z$ is sufficiently large, or the electron has been traveling in the inhomogeneous magnetic field long enough such that $m$ can be approximated as infinity. Since $\theta\in [0, \pi]$ and $\cos(\theta) \in [-1, 1]$, we have
\begin{equation}
        \lim_{m\to\infty}[\cos(\theta)]^{2m}  = \left\{ \begin{array}{rcl}  0, & \mbox{for} & \theta\in(0, \pi) \\
         1, & \mbox{for} & \theta = 0, \pi \end{array}\right.
\end{equation}
Thus, the probability density function is
\begin{equation}
\label{pinfty}
        \lim_{m\to\infty}p_m(\theta)  \propto \left\{ \begin{array}{rcl}  0, & \mbox{for} & \theta\in(0, \pi) \\
         1, & \mbox{for} & \theta = 0, \pi \end{array}\right.
\end{equation}
This shows that when $\nabla B_z$ is sufficiently large, or the electron has been traveling in the inhomogeneous magnetic field for a distance long enough, measurement of the electron spin can only obtain two discrete outcomes: spin up and spin down, along the $z$-axis, as shown in Figure 2. This explains the quantization of spin measurement outcomes. 

However, (\ref{pinfty}) itself cannot be a valid probability density function. Its integral is zero since it is only nonzero when $\theta = 0, \pi$. The Dirac delta function is the proper mathematical tool to describe such a situation. We can rewrite
\begin{equation}
\label{pinfty2}
   \Bar{p}(\theta) := \lim_{m\to\infty}p_m(\theta) = \frac{1}{2}\{\delta(\theta) + \delta(\theta-\pi)\}.
\end{equation}
The factor of $1/2$ is due to two facts, 1.) the normalization requirement, and 2.) the result of the spin-up or spin-down measurement is completely random. The second point here needs more elaboration. Before applying the external inhomogeneous field $B_z$, the intrinsic angular momentum of the electron is completely randomly oriented. At the moment when $B_z$ is applied, it is half the chance that the initial direction of the angular momentum forms an angle with the $z$-axis $\theta$ such that $\theta\in [0, \pi/2]$. As the electron travels along $B_z$, the angular momentum vector evolves closer and closer to the $z$ -axis and eventually spins up. With the other half of chance, the initial direction of the angular momentum forms an angle with the $z$-axis $\theta$ such that $\theta\in [\pi/2, \pi]$. As the electron travels along $B_z$, the angular momentum evolves to be spin-down. For simpler notation, define
\begin{equation}
    \Theta^+ := \{\forall \theta \in [0, \pi/2]\},  \mbox{ and } \Theta^-:= \{\forall \theta \in [\pi/2, \pi]\}.
\end{equation}
Then, $\theta \in \Theta^+$ means that the spin orientation is pointing to the upper half of the orientation sphere. Similarly, $\theta\in\Theta^-$ means that the spin orientation is pointing to the lower half of the orientation sphere. The factor $1/2$ is therefore due to the complete initial randomness of the orientation of the intrinsic angular momentum such that the probability of $\theta\in\Theta^+$ is the same as the probability of $\theta\in\Theta^-$. 
\begin{figure}
\begin{center}
\includegraphics[scale=0.7]{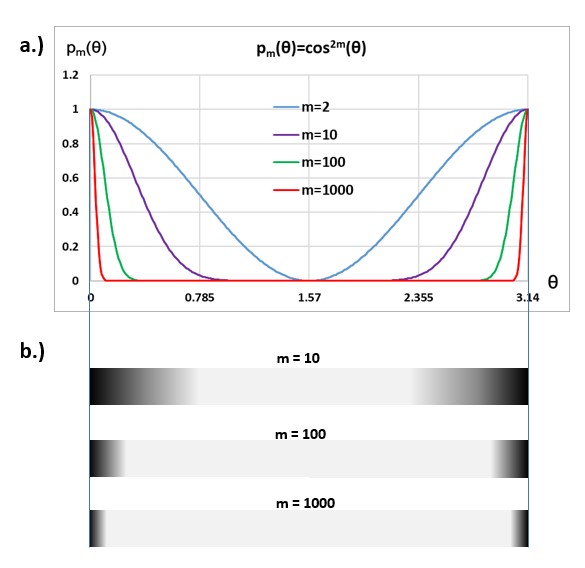}
\caption{a.) As parameter $m$ increases, the probability distribution converges to only two directions at $\theta=0$ and $\theta=\pi$. b.) The envisioned Stern-Garlach measurement results with an inhomogeneous magnetic field. As the electrons travel along $B_z$, the electrons reach the detectors that eventually show two converging discrete lines.}
\end{center}
\label{fig2}
\end{figure}

The above analysis implies that if the initial probability density before measurement is $\sigma (\theta)$, the probability of measurement outcome as spin-up can be calculated as
\begin{equation}
\label{pup}
    p(\uparrow) := \varrho(\Theta^+) = \int_0^{\pi/2} \sigma (\theta) d\theta.
\end{equation}
Similarly, the probability of measurement outcome as spin-down is
\begin{equation}
\label{pdown}
    p(\downarrow) := \varrho(\Theta^-) = \int_{\pi/2}^{\pi} \sigma (\theta) d\theta,
\end{equation}
and $\varrho(\Theta^+)+\varrho(\Theta^-)=1$. The generalized form of (\ref{pinfty2}) is
\begin{equation}
\label{pinfty3}
   \Bar{p}(\theta) = \varrho(\Theta^+)\delta(\theta) + \varrho(\Theta^-)\delta(\theta-\pi).
\end{equation}
We will derive \eqref{pinfty3} more rigorously in later sections. For an initially uniformly distributed orientation, $\sigma$ is a constant, and we get $p(\uparrow) = p(\downarrow) = 1/2$.

The probability of $1/2$ for spin-up and spin-down is due to the complete randomness of the initial orientations of the intrinsic angular momentum. It is important to note that the measurement results are relative to the direction defined by the magnetic field $B_z$. Given the same initial random orientations, if one measures the spin along a different direction, say $z'$-axis, one will still get either spin-up and spin-down with probability of $1/2$. However, if the initial orientation of the intrinsic angular momentum is not completely random, the probability of obtaining spin-up and spin-down along the $z'$-axis will be different from 1/2. This is the subject of study in the next section.

\begin{figure*}
\begin{center}
\includegraphics[scale=0.6]{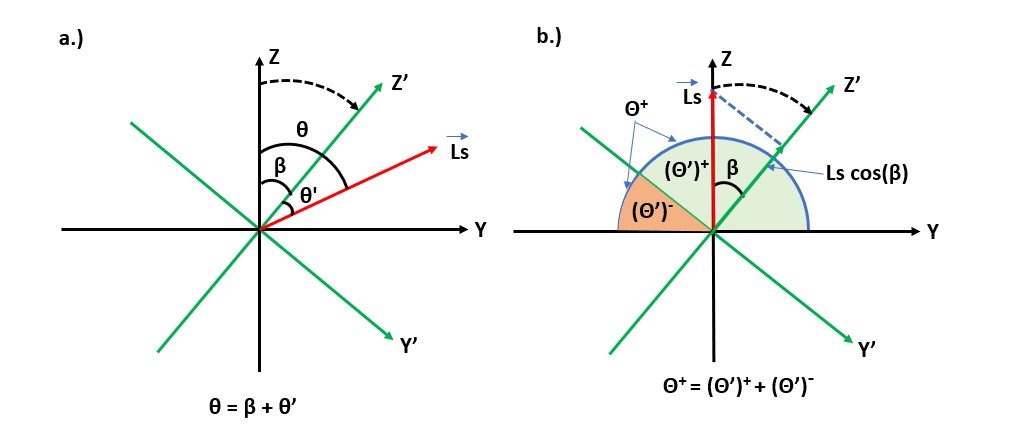}
\caption{a.) When the direction of inhomogeneous magnetic field in the second Stern-Gerlach apparatus is rotated with an angle $\beta$ with the respect to the $z$-axis, the orientation of intrinsic angular momentum is given by $\theta'=\theta+\beta$. b.) The range of upward angle with respect to the $z$-axis, $\Theta^+$, is split into two parts, $(\Theta')^+$, and $(\Theta')^-$, with respect to the $z'$-axis. Thus, the second Stern-Gerlach apparatus gives both possible measurement outcomes of spin-up and spin-down.}
\end{center}
\label{fig3}
\end{figure*}

\subsection{Subsequent Measurement with a Rotated Stern-Gerlach Apparatus}
Suppose after the electron passes through a Stern-Gerlach apparatus and is observed the measurement result is spin-up. Now, letting the electron pass a subsequent Stern-Gerlach apparatus with the exact same direction of inhomogeneous magnetic field, one will obtain the result of spin-up for the electron. This means that the electron remains in the spin-up state after passing through the first Stern-Gerlach apparatus\footnote{In standard quantum mechanics, this is explained as that the electron stays in the eigen-state of spin-up.}. However, in our model, after the electron passes through the first Stern-Gerlach apparatus and the external magnetic field is removed, due to the nature of random fluctuations, the orientation of the intrinsic electron angular momentum can start to deviate from the $z$-axis with an angle $\theta$ between the angular momentum and the $z$-axis. In other words, due to fluctuations, the orientation of the angular momentum is relaxing away from pointing along the $z$-axis. This process is important in order to explain the behavior if the second Stern-Gerlach apparatus is set up such that the direction of the magnetic field is tilted with an angle $\beta$ from the original $z$-axis. Denote this new direction by $z'$-axis and the magnetic field by $B_{z'}$.

Let the probability density conditioned on the initial measurement outcome of spin-up be $p(\theta'|\theta)$, where $\theta'$ is the angle between the orientation of the intrinsic angular momentum of the electron and the $z'$-axis. From Figure 3, it can be seen that $\theta = \theta'+ \beta$. We will apply the extended least action principle again to find out $p(\theta'|\theta)$. With the magnetic field $B_{z'}$, the classical action can be calculated similarly to that in (\ref{spinAction}) - (\ref{spinAction2}) as
\begin{equation}
\label{spinAction3}
    A'_c = -\frac{1}{2}g_sL_s\int_0^{\pi}\int_0^{\Delta \varphi} p(\theta'|\theta) \cos(\theta')d\theta' d\varphi.
\end{equation}
To compute $I_f$, we choose the Tsallis divergence again. For the infinitesimal period of time $\Delta t$, we define
\begin{equation}
    I_f = \frac{1}{\alpha-1}\{\int_0^{\pi}\int_0^{\Delta \varphi} \frac{p^{\alpha}(\theta'|\theta)}{\sigma^{\alpha-1}(\theta)}d\theta' d\varphi - 1 \},
\end{equation}
where $\sigma(\theta)$ is no longer an uniform probability density since the initial condition is that the electron has passed through the first Stern-Gerlach apparatus. Applying the same variation principle as that to derive (\ref{ptheta}) and (\ref{pthetam}), we obtain
\begin{equation}
    \label{pthetaprime}
    p_m(\theta'|\theta)=\frac{1}{Z_m}\sigma(\theta)[\cos(\theta')]^{2m}, m \in \mathbb{N}.
\end{equation}
When $\nabla B_{z'}$ is sufficiently large, or the electron travels sufficiently long distance in the magnetic field, we get
\begin{equation}
\label{pinftyprime}
   \bar{p}(\theta'|\theta) := \lim_{m\to\infty}p_m(\theta'|\theta) =\sigma(\theta)\{\delta(\theta') + \delta(\theta'-\pi)\},
\end{equation}
where the normalization factor is omitted. Noting that $\theta = \theta'+ \beta$, we can compute the probability of obtaining measurement of spin-up along the $z'$-axis as
\begin{equation}
\label{pupprime}
    p(\uparrow|\beta) \propto \int_0^{\pi/2} \bar{p}(\theta'|\theta) d\theta' = \sigma(\beta).
\end{equation}
Similarly, the probability of obtaining spin-down is
\begin{equation}
\label{pupprime}
    p(\downarrow|\beta) \propto \int_{\pi/2}^{\pi} \bar{p}(\theta'|\theta) d\theta' = \sigma(\beta+\pi).
\end{equation}
To impose the normalization requirement, we must have $p(\uparrow|\beta)+p(\downarrow|\beta)=1$. Denote $Z:=\sigma(\beta)+\sigma(\beta+\pi)$, the normalization can be achieved by setting 
\begin{align}
    \label{normalpprime}
    p(\uparrow|\beta) &= \sigma(\beta) / Z := \varrho(\beta) \\
    p(\downarrow|\beta) &= \sigma(\beta+\pi) / Z := \varrho(\beta+\pi)
\end{align}
Then, \eqref{pinftyprime} can be rewritten as
\begin{equation}
\label{pinftyprime2}
    \bar{p}(\theta'|\beta) =\varrho(\beta)\delta(\theta') + \varrho(\beta+\pi)\delta(\theta'-\pi).
\end{equation}
Notice that \eqref{pinftyprime2} is the same as \eqref{pinfty3} if we set $\varrho (\beta) = \varrho(\Theta^+)$ and $\varrho(\beta+\pi) = \varrho(\Theta^-)$. This effectively proves \eqref{pinfty3}.

The average angular momentum along the $z'$-axis will be $(\varrho(\beta) -\varrho(\beta+\pi))L_s$. On the other hand, from Figure 3, one can deduce that this average angular momentum must be $L_s\cos(\beta)$, thus\footnote{If the initial condition is spin-down after the electron passing the first Stern-Gerlach apparatus, the average angular momentum becomes $-L_s\cos(\beta)$, resulting $\varrho(\beta) - \varrho(\beta+\pi) = -\cos(\beta)$.}
\begin{equation}
    \label{averagedL}
    \varrho(\beta) - \varrho(\beta+\pi) = \cos(\beta).
\end{equation}
Together with the normalization condition $\varrho(\beta) + \varrho(\beta+\pi)=1$, we obtain
\begin{equation}
    \label{averagedL2}
    \left\{ \begin{array}{lcl}  \varrho(\beta) = (1+\cos(\beta))/2 =\cos^2(\beta/2)\\
         \varrho(\beta+\pi) =(1 - \cos(\beta))/2 = \sin^2(\beta/2)\end{array}\right.,
\end{equation}
Therefore, we reproduce the same result as that from the standard quantum mechanics,
\begin{equation}
    \label{pupbeta}
    \left\{ \begin{array}{lcl}  p(\uparrow|\beta) =\cos^2(\beta/2) \\
         p(\downarrow|\beta) = \sin^2(\beta/2)\end{array}\right..
\end{equation}
Eq.\eqref{pinftyprime2} now becomes
\begin{equation}
\label{pinftyprime3}
    \bar{p}(\theta'|\beta) =\cos^2(\beta/2)\delta(\theta') + \sin^2(\beta/2)\delta(\theta'-\pi).
\end{equation}

Suppose that $\beta < \pi/2$. We wonder why there is the possibility that the measurement result of the second Stern-Gerlach apparatus is spin-down. That is, why $p(\downarrow|\beta)$ is non-zero? The question arises because if the initial condition is spin-up after the electron passes through the first Stern-Gerlach apparatus, and if the orientation of the intrinsic angular momentum is kept unchanged, the initial angle $\theta'=\beta \in (0, \pi/2)$. Then, according to the discussions leading to (\ref{pup}), we should only obtain spin-up as the only result when $B_{z'}$ is applied. However, as pointed out at the beginning of this section, after the electron passes through the first Stern-Gerlach apparatus, the direction of the intrinsic angular momentum starts to deviate from the $z$-axis due to vacuum fluctuations. Suppose after a sufficient relaxation time $\tau_r$, the orientation of the intrinsic angular momentum of the electron can be at any angle $\theta$ from the $z$-axis with $\theta \in [0, \pi/2]$. When the electron enters the second Stern-Gerlach with the magnetic field $B_{z'}$ being tilted at an angle $\beta \in (0, \pi/2)$, the initial orientation is now measured with respect to the $z'$-axis as angle $\theta'$. Since $\theta'=\theta-\beta$, the orientation of the intrinsic angular momentum can point to either upward with $\theta'\in (\Theta')^+$, or downward with $\theta'\in(\Theta')^-$, as shown in Figure 3b. Thus, there is a possibility of spin-down when $\nabla B_{z'}$ is sufficiently large. We see that relaxation of the orientation of intrinsic angular momentum is necessary to explain the result in (\ref{pupbeta}), which has been confirmed experimentally.

There is still a puzzle to be clarified. The second Stern-Gerlach apparatus can be rotated so that $\theta - \theta'$ is $\beta$ or $-\beta$, the probability of measurement outcome for spin-up or spin-down \eqref{pinftyprime3} is the same for both cases. This suggests that \eqref{pinftyprime3} does not completely specify the behavior of the electron spin in a single measurement. To remedy that, we need to specify the probability of measurement outcome along another axis, say the $y$-axis. Suppose the $y-z$ plane is rotated along the $x$-axis such that $z'$-axis is deviated from $z$-axis by an angle $\beta$, then the angle between $y'$-axis and $z$-axis is $\pi/2+\beta$. The probability density of measurement outcome along the $y'$-axis can be obtained by replacing parameter $\beta$ in \eqref{pinftyprime3} with $\pi/2+\beta$,
\begin{equation}
\label{py}
\begin{split}
    \bar{p}(\theta'_y|\frac{\pi}{2}+\beta) &=\frac{1}{2}[\cos(\frac{\beta}{2})-\sin(\frac{\beta}{2})]^2\delta(\theta'_y) \\
    & + \frac{1}{2}[\cos(\frac{\beta}{2})+\sin(\frac{\beta}{2})]^2\delta(\theta'_y-\pi).
\end{split}
\end{equation}
On the other hand, if the $z'$-axis is deviated from the $z$-axis by an angle $-\beta$, then the angle between the $y'$-axis and the $z$-axis is $\pi/2-\beta$. The probability density of measurement outcome along the $y'$-axis is
\begin{equation}
\label{py2}
\begin{split}
    \bar{p}(\theta'_y|\frac{\pi}{2}-\beta) &=\frac{1}{2}[\cos(\frac{\beta}{2})+\sin(\frac{\beta}{2})]^2\delta(\theta'_y) \\
    & + \frac{1}{2}[\cos(\frac{\beta}{2})-\sin(\frac{\beta}{2})]^2\delta(\theta'_y-\pi).
\end{split}
\end{equation}
It is clear that \eqref{py} and \eqref{py2} are different. Thus, \eqref{pinftyprime3} and \eqref{py} together give a complete description of spin behavior when the second Stern-Gerlach is rotated by an angle $\beta$, while \eqref{pinftyprime3} and \eqref{py2} together give a complete description of spin behavior when the second Stern-Gerlach is rotated by an angle $-\beta$.

We can generalize \eqref{pupbeta} by assuming that the magnetic field of the first Stern-Gerlach apparatus is along a direction tilted at an angle $\beta_1$ from the $z$-axis, and the second Stern-Gerlach apparatus is along a direction tilted at an angle $\beta_2$ from the $z$-axis. Suppose that measurement outcome of the first Stern-Gerlach apparatus is spin-up, what is the probability of measurement outcome with spin-up from the second Stern-Gerlach apparatus? The calculation steps from (\ref{spinAction3}) to (\ref{pupbeta}) can be repeated but with $\theta = \theta'+ \beta_2 - \beta_1$. Thus, we can just replace $\beta$ with $\beta_2-\beta_1$ in steps from (\ref{spinAction3}) to (\ref{pupbeta}), and the final result will be
\begin{equation}
    \label{pupbeta2}
    \left\{ \begin{array}{lcl}  p(\uparrow|\beta_2, \beta_1) =\cos^2((\beta_2-\beta_1)/2) \\
         p(\downarrow|\beta_2, \beta_1) = \sin^2((\beta_2-\beta_1)/2)\end{array}\right..
\end{equation}
In Appendix \ref{appendix:rotation}, it is verified that (\ref{pupbeta2}) is exactly the same result predicted by standard quantum mechanics.

\section{Derivation of the Schr\"{o}dinger-Pauli Equation}
\label{sec:PauliEq}
The framework to derive the law of dynamics for an electron in a external magnetic field $B_z$, for a cumulative period from $t_a$ to $t_b$, is similar to that described in Appendix \ref{sec:shorttime}, except that we need to add a new term in the Lagrangian due to the interaction of spin and the external magnetic field. Based on the results shown in (\ref{pinfty2}) and (\ref{pinftyprime2}), the probability density can be written as
\begin{equation}
    p(\theta) = \sum_i \sigma_i\delta(\theta-\theta_i).
\end{equation}
For an electron, we only need to consider the two-level case where $i\in \{0,1\}$ and $\theta_0=0, \theta_1=\pi$. The expectation value of the potential energy due to the the interaction of spin and $B_z$ is
\begin{equation}
    \label{averagedU}
    \begin{split}
    \overline{U} &= -\int p(\theta) \vec{\mu}\cdot \Vec{B}  d\theta \\
            &= \frac{e\hbar}{2m}\int p(\theta) B_z\cos(\theta) d\theta,
    \end{split}
\end{equation}
where in the second step, we substitute the magnetic moment $\mu$ with \eqref{muLs} and choose the g-factor $g_s=2$. 

More generically, we need to specify the probability density with the space-time coordinates. Denote 
\begin{equation}
    \label{pgeneral}
    \rho(\mathbf{x}, t, \theta) = \sum_i \sigma_i(\mathbf{x}, t, \theta_i)\delta(\theta-\theta_i).
\end{equation}
Then the expectation value of the potential energy is 
\begin{equation}
    \label{averagedU2}
    \begin{split}
        \overline{U} &= \frac{e\hbar}{2m}\int \rho(\mathbf{x}, t, \theta) B_z \cos(\theta) d\theta d^3\mathbf{x} \\
        & = \frac{e\hbar}{2m}\sum_i \int\sigma_i(\mathbf{x}, t, \theta_i)\delta(\theta-\theta_i)B_z \cos(\theta) d\theta d^3\mathbf{x}.
    \end{split}    
\end{equation}
In the case of electrons, there are only two values $\theta_0=0, \theta_1=\pi$. Therefore,
\begin{equation}
    \label{averagedU3}
    \begin{split}
        \overline{U} & = \frac{e\hbar}{2m}\{\int\sigma_0(\mathbf{x}, t, 0)\delta(\theta)\ B_z \cos(\theta) d\theta d^3\mathbf{x} \\
        &+ \int\sigma_1(\mathbf{x}, t, \pi)\delta(\theta-\pi) B_z \cos(\theta) d\theta d^3\mathbf{x} \} \\
        & = \frac{e\hbar}{2m}\{\int\sigma_0(\mathbf{x}, t, 0) B_z d^3\mathbf{x} -  \int\sigma_1(\mathbf{x}, t, \pi) B_z d^3\mathbf{x} \}.
    \end{split}    
\end{equation}
Here we run into a problem. In the Stern-Gerlach experiment, it is confirmed that electrons with spin-up follow a different trajectory from that of electrons with spin-down. An electron with either spin-up or spin-down follows separated laws of dynamics. We cannot simply add the two terms in (\ref{averagedU3}) together. Instead, a more proper notation should treat the probability density as a two-component vector. Thus,
\begin{equation}
    \label{averagedU4}
    \overline{U}  = \frac{e\hbar}{2m}\int B_z \left( \begin{array}{c} \sigma_0(\mathbf{x}, t, 0) \\ -\sigma_1(\mathbf{x}, t, \pi) \end{array} \right) dx := \left( \begin{array}{c} \overline{U}_+ \\ \overline{U}_- \end{array} \right).
\end{equation}
For further simplification of notation, denote $\rho_+(x,t)=\sigma_0(x, t, 0)$ and \added{$\rho_-(x,t)=\sigma_1(x, t, \pi)$}. They should satisfy the normalization condition
\begin{equation}
    \label{normalPM}
    \int [\rho_+(\mathbf{x},t) + \rho_-(\mathbf{x},t)]dx = 1.
\end{equation}

First we will derive the law of dynamics for the electron with spin up. Without considering the contribution from the spin magnetic interaction, the classical action of an electron in an external electromagnetic field described by the magnetic vector potential\footnote{Note that $\vec{B} = \nabla\times\Vec{A}$. Since we only consider the case where $\vec{B}$ is along the $z$-axis, $\nabla\times\Vec{A}$ only has the $z$ component. } $\Vec{A}$ and electric scalar potential $\phi$ is given by (see Appendix C of \cite{Yang2023})
\begin{equation}
    \label{cAction2}
    A_c = \int\rho_+\{ \frac{\partial S_+}{\partial t} + \frac{1}{2m}(\nabla S_+ + e\Vec{A})^2 - e\phi\} d^3\mathbf{x}dt
\end{equation}
Now we need to add the additional term of potential energy $\overline{U}_+$ due to spin magnetic interaction into \eqref{cAction2}
\begin{equation}
    \label{cAction3}
    A_c^+ = \int\rho_+\{ \frac{\partial S_+}{\partial t} + \frac{1}{2m}(\nabla S_+ + e\Vec{A})^2 +\frac{e\hbar}{2m} B_z - e\phi\} d^3\mathbf{x}dt
\end{equation}
The definition of information metrics for the vacuum translational fluctuation is the same as \eqref{DLDivergence}, and similar to the calculation shown in \cite{Yang2023}, when $\Delta t\to 0$ it becomes
\begin{equation}
\label{FisherInfoPlus}
    I_f^+ = \int d^3\mathbf{x}dt \frac{\hbar}{4m}\frac{1}{\rho_+}\nabla\rho_+ \cdot \nabla\rho_+.
\end{equation}
Inserting \eqref{cAction3} - \eqref{FisherInfoPlus} into \eqref{totalAction}, we have
\begin{equation}
    \label{totalActionSpin}
    \begin{split}
        A_t^+ =& \int\rho_+\{ \frac{\partial S_+}{\partial t} + \frac{1}{2m}(\nabla S_+ + e\Vec{A})^2 +\frac{e\hbar}{2m} B_z \\
        &- e\phi\} d^3\mathbf{x}dt  + \frac{\hbar^2}{8m} \int \frac{1}{\rho_+}\nabla\rho_+ \cdot \nabla\rho_+ d^3\mathbf{x}dt.
    \end{split}   
\end{equation}
Performing the variation procedure on $A_t$ with respect to $S_+$ gives
\begin{equation}
\label{contPlus}
    \frac{\partial\rho_+}{\partial t }+ \frac{1}{m}\nabla \cdot(\rho_+ (\nabla S_+ + e\Vec{A})) = 0,
\end{equation}
which is the continuity equation for $\rho_+$. Performing the variation procedure on $A_t$ with respect to $\rho_+$ leads to the spin-up version of extended Hamilton-Jacobi equation
\begin{equation}
\label{QHJplus}
    \frac{\partial S_+}{\partial t} + \frac{1}{2m}(\nabla S_+ + e\Vec{A})^2 +\frac{e\hbar}{2m} B_z - e\phi - \frac{\hbar^2}{2m}\frac{\nabla^2\sqrt{\rho_+}}{\sqrt{\rho_+}} = 0.
\end{equation}
Defined a complex function $\Psi_+=\sqrt{\rho_+}e^{iS_+/\hbar}$, the continuity equation \eqref{contPlus} and the extended Hamilton-Jacobi equation \eqref{QHJplus} can be combined into a single differential equation,
\begin{equation}
    \label{SEup}
    i\hbar\frac{\partial\Psi_+}{\partial t} = [\frac{1}{2m}(i\hbar\nabla + e\Vec{A})^2 + \frac{e\hbar}{2m} B_z - e\phi]\Psi_+,
\end{equation}
which is the Schr\"{o}dinger equation for a spin-up electron. 

The derivation of the dynamics equations for a spin-down electron is exactly the same, except there is a sign difference for the potential energy term $\overline{U}_-$. The resulting Schr\"{o}dinger equation is 
\begin{equation}
    \label{SEdown}
    i\hbar\frac{\partial\Psi_-}{\partial t} = [\frac{1}{2m}(i\hbar\nabla + e\Vec{A})^2 -\frac{e\hbar}{2m} B_z - e\phi]\Psi_-,
\end{equation}
where $\Psi_-=\sqrt{\rho_-}e^{iS_-/\hbar}$.

To combine equations \eqref{SEup} and \eqref{SEdown}, we introduce a two-component vector wave function
\begin{equation}
    \label{vectorWF}
    \Psi = \left( \begin{array}{c} \Psi_+ \\ \Psi_- \end{array} \right)
\end{equation}
and a two dimensional matrix
\begin{equation}
    \label{vectorWF}
    \sigma_z = \left( \begin{array}{cc} 1 & 0 \\ 0 & -1 \end{array} \right),
\end{equation}
then \eqref{SEup} and \eqref{SEdown} can be combined into a compact form
\begin{equation}
    \label{PauliE}
    i\hbar\frac{\partial\Psi}{\partial t} = [\frac{1}{2m}(i\hbar\nabla + e\Vec{A})^2 + \frac{e\hbar}{2m} \sigma_z B_z -e\phi]\Psi,
\end{equation}
which is the Schr\"{o}dinger-Pauli equation for an electron in an external magnetic field along the $z$-axis $B_z$. The normalization condition \eqref{normalPM} is rewritten as
\begin{equation}
    \label{normalPsi}
    \int \Psi^+\Psi dx = \int (\Psi^*_+\Psi_+ + \Psi^*_-\Psi_-) d^3\mathbf{x} = 1.
\end{equation}

Generalizing the Schr\"{o}dinger-Pauli equation for an electron in an external magnetic field $\Vec{B}$ along arbitrary orientation is the natural next step, as it will demonstrate the rotational property of spin. We will leave it for future research.

\begin{figure*}
\begin{center}
\includegraphics[scale=0.75]{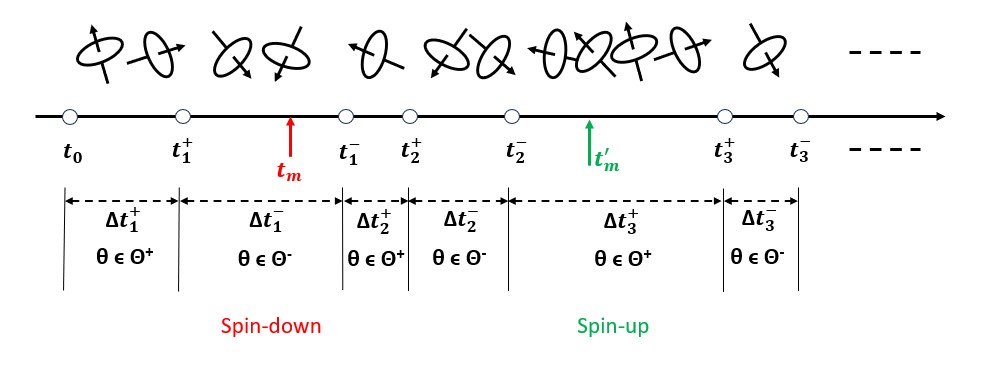}
\caption{Randomness of orientation of intrinsic angular momentum from time perspective. If the measurement occurs at a time $t_m$ that the initial orientation is $\theta\in\Theta^-$, the measurement outcome is spin-down. On the other hand, if the measurement occurs at a time $t'_m$ that the initial orientation $\theta\in\Theta^+$, the measurement outcome is spin-up}
\end{center}
\label{fig4}
\end{figure*}

\section{Entanglement}
\label{sec:entanglement}
\subsection{Spin Correlation}
\label{sec:SpinEnt}
To investigate the entanglement phenomenon between two electron spins, we need to take a closer look at the time dynamics of the fluctuations of spin orientations. As shown earlier, when measuring the spin of an electron using an inhomogeneous external magnetic field along any axis, one will obtain two discrete results, spin-up or spin-down along the axis. Assign this axis as the $z$-axis. Whether the result is spin-up or spin-down is random, depending on the initial orientation relative to the $z$-axis, described by the angle $\theta$. To calculate the probability of the initial orientation for $\theta \in \Theta^+$ or $\theta \in \Theta^-$, we need to analyze how the orientation of intrinsic angular momentum fluctuates before an external magnetic field $B_z$ is applied. Recall that we denote $\theta \in \Theta^+$ as the orientation of the angular momentum that is trending upward with angle $\theta\in [0,\pi/2]$, and $\theta \in \Theta^-$ as the orientation that is trending downward. The amount of time for the orientation to change from $\theta \in \Theta^+$ to $\theta \in \Theta^-$ must be non-zero. That is, it takes a non-zero amount of time for the orientation of intrinsic angular momentum to change from trending upward to trending downward, and vice versa. 
Suppose that at $t_0$, $\theta \in \Theta^+$. At time $t_1^+$, $\theta$ changes to a downward trend, $\theta \in \Theta^-$. So, the angular momentum continues to trend upward for a period of time $\Delta t^+_1 = t_1^+ - t_0$. Then the orientation stays trending downward for a duration $\Delta t^-_1$ until at time $t_1^-$, it moves back to pointing upward. After the orientation continues to trend upward for a period of time $\Delta t^+_2$, it changes to point downward for a period of time $\Delta t^-_2$. The process of switching orientation directions continues to $t=\Delta T$, as shown in Figure 4. Statistically, $\{\Delta t^+_i, i \in \mathbb{N}\}$ forms a random variable, and \replaced {so as for $\{\Delta t^-_i, i \in \mathbb{N}\}$}{similarly for $\{\Delta t^+_-, i \in \mathbb{N}\}$}. Define
\begin{equation}
    \Delta T^+ = \sum_i \Delta t_i^+, \mbox{ and } \Delta T^- = \sum_i \Delta t_i^-,
\end{equation}
we have $\Delta T^+ + \Delta T^- = \Delta T$. Suppose that a spin measurement experiment is conducted at $t_m$. If $t_m$ occurs at the moment when $\theta \in \Theta^+$, that is, $t_m$ falls into one of the time periods $\Delta t^+_i$, as shown in Figure 4, the measurement result will be spin-up. On the other hand, if $t_m$ falls into one of the time periods $\Delta t^-_i$, the measurement result will be spin-down. Since $\Delta t^+_i$ and $\Delta t^-_i$ are random variables, the measurement outcome are random as well, and the probabilities of orientation trending upward and downward are
\begin{equation}
    \label{ptime}
        \varrho(\Theta^+) = \lim_{\Delta T \to \infty}\frac{\Delta T^+}{\Delta T}, \mbox{ and } \varrho(\Theta^-) = \lim_{\Delta T \to \infty}\frac{\Delta T^-}{\Delta T},
\end{equation}
respectively\footnote{Eq. \eqref{ptime} can be considered as the frequency interpretation of $\varrho (\cdot)$ in Eqs. \eqref{pup} and \eqref{pdown}, while Eqs. \eqref{pup} and \eqref{pdown} are the classical interpretation. Both interpretations are considered equivalent.}. For an intrinsic angular momentum with completely random orientations, we must have $\Delta T^+ = \Delta T^-$, and therefore $\varrho(\Theta^+) =\varrho(\Theta^-) = 1/2$.

The random variable $\{\Delta t^+_i \ge 0, i \in \mathbb{N}\}$ can follow a probability distribution. Here we will not investigate the actual probability distribution. What is relevant to our investigation is the expectation value, denoted as
\begin{equation}
    \label{meanT}
    \tau^+ = \langle \Delta t^+_i \rangle, \mbox{ and similarly, } \tau^- = \langle \Delta t^-_i \rangle.
\end{equation}
Again, for an intrinsic angular momentum with completely random orientation, it is intuitive to assume $\tau^+ = \tau^-$.

In summary, the spin model of an electron developed here has the following characteristics:
\begin{itemize}
    \item Applying an external inhomogeneous magnetic field results in two discrete measurement outcomes, spin-up or spin-down;
    \item The statistical probability of obtaining spin-up or spin-down is determined by the initial probability $\varrho(\Theta^+)$ and $\varrho(\Theta^-)$;
    \item For a particular measurement event occurred at $t_m$, whether the outcome is spin-up or spin-down depends on $\theta \in \Theta^+$ or $\theta \in \Theta^-$ at $t_m$, respectively.
\end{itemize}

Now we consider the case of two electrons $A$ and $B$. Suppose that in the preparation stage of experiment setup they together go through interactions with a common source of external field until $t_0$. As a result, they share some kinds of correlation even though there is no external field applied to them after $t_0$. Consider a particular type of correlation between these two electrons such that their intrinsic angular momenta always point to the opposite orientations. This implies that $\theta_A + \theta_B = \pi$ at any time and that the reference axis can be along any direction until a measurement event occurs. The correlation $\theta_A + \theta_B = \pi$ is maintained even though the orientations of both intrinsic angular momenta fluctuate randomly. Consequently, we have $\varrho(\Theta^+_A)=\varrho(\Theta^-_B)$, $\varrho(\Theta^-_A)=\varrho(\Theta^+_B)$, and $\tau^+_A=\tau^-_B$, $\tau^-_A=\tau^+_B$. 

Next, we want to see what happens when a measurement event occurs. There are two cases here; measurements on $A$ and $B$ occur at the same time or at different times. First, suppose that the two electrons are measured at the same time, which is the usual experiment setup for a typical Bell test. For electron $A$, the probability of measurement outcome can be described as
\begin{equation}
    \label{measA}
    \Tilde{p}(\theta_A) = \varrho(\Theta^+_A)\delta(\theta_A) + \varrho(\Theta^-_A)\delta(\theta_A-\pi).
\end{equation}
For electron $B$, due to the correlation $\theta_B + \theta_A = \pi$, the measurement results will be exactly anti-correlated to the results of $A$. For example, if the measurement result of $A$ is spin-up, it implies the initial condition $\theta_A \in \Theta^+_A$, which means that the orientation of the angular momentum is trending upward with any angle
$\theta_A \in [0, \pi/2)$ relative to the $z$-axis. Since $\theta_B + \theta_A = \pi$, $\theta_B$ must also trend downward relative to the same direction. Thus, the measurement outcome must be spin-down. Similarly, if the measurement result of $A$ is spin-down, then the measurement result of $B$ must be spin-up. However, note that the measurement result of $A$ is random, depending on whether the initial condition $\theta_A$ is trending upward or downward relative to the $z$-axis at $t_m$. The correlation of measurement results can be described by a joint probability density
\begin{equation}
    \label{measAB2}
    \Tilde{p}(\theta_A, \theta_B) = \varrho(\Theta^+_A)\delta(\theta_A)\delta(\theta_B-\pi) + \varrho(\Theta^-_A)\delta(\theta_A-\pi)\delta(\theta_B).
\end{equation}
The subscript $A$ can be removed in the expression $\varrho(\Theta^+_A)$ since $\varrho(\Theta^+_A)=\varrho(\Theta^-_B)$. In the case that the initial orientations of angular momentum of electron $A$ are uniformly distributed for all $\theta\in[0,\pi]$, one has $\varrho(\Theta^+_A)=\varrho(\Theta^-_A)=1/2$, \eqref{measAB2} becomes
\begin{align}
    \label{measABz}
    \Tilde{p}(\theta_A^z, \theta_B^z) = \frac{1}{2}\delta(\theta_A^z)\delta(\theta_B^z-\pi) + \frac{1}{2}\delta(\theta_A^z-\pi)\delta(\theta_B^z),
\end{align}
where the superscript $z$ is added to explicitly show the context dependency on the $z$-axis. However, Eq. \eqref{measABz} itself does not uniquely specify the effect of correlation when the angular momentum orientations of the electron pair are always opposite to each other, because \eqref{measABz} can correspond to two possible correlations, as shown in Figure 5. To uniquely specify the effect of correlation when the angular momentum orientations of the electron pair are always opposite to each other, one needs to show the probability density along another axis perpendicular $z$-axis, say the $y$-axis. Consider the case that the correlations between the electron pair are such that not only $\theta_A^z+\theta^z_B=\pi$ along the $z$-axis, but also $\theta_A^y+\theta^y_B=\pi$ along the $y$-axis. This implies that we must also have
\begin{equation}
    \label{measABy}
    \Tilde{p}(\theta_A^y, \theta_B^y) = \frac{1}{2}\delta(\theta_A^y)\delta(\theta_B^y-\pi) + \frac{1}{2}\delta(\theta_A^y-\pi)\delta(\theta_B^y).
\end{equation}

\begin{figure}
\begin{center}
\includegraphics[scale=0.6]{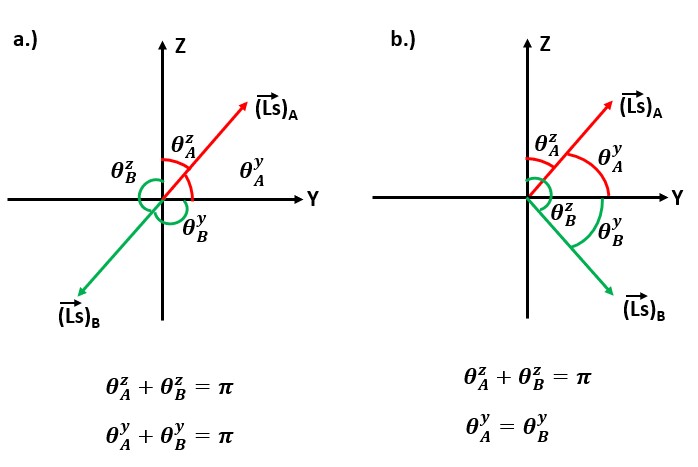}
\caption{Correlation of intrinsic angular momenta between two electrons. a.) Correlation corresponds to the singlet state $|\Psi^-\rangle$; b.) Correlation corresponds to Bell state $|\Psi^+\rangle$. Note that for both a.) and b.), we have $\theta_A^z+\theta^z_B=\pi$. It shows $\theta_A^z+\theta^z_B=\pi$ alone is unable to distinguish the correlations between a.) and b.). Thus, we need to also specify the correlation along the $y$-axis.}
\end{center}
\label{fig5}
\end{figure}

When a pair of electrons is described by both \eqref{measABz} and \eqref{measABy}, the correlation between the two electrons is equivalent to that described by the spin singlet state in standard quantum mechanics.
\begin{equation}
    \label{singlet}
    |\Psi^-\rangle := \frac{1}{\sqrt{2}}(|\uparrow\rangle_A|\downarrow\rangle_B - |\downarrow\rangle_A|\uparrow\rangle_B).
\end{equation}
The equivalence of \eqref{singlet} versus \eqref{measABz} and \eqref{measABy} will be verified in the next section when we calculate the Bell-CHSH inequality. 

Before considering the more complicated case where measurements of $A$ and $B$ occur at different times, we need to \replaced{clarify two important questions. First,}{examine} how is the correlation $\theta_B + \theta_A = \pi$ maintained when the two electrons are separated? Recall that in our spin model, the intrinsic angular momentum of the electron is not due to rotation as a rigid body, but due to the rotational components of vacuum fluctuations of the electrons, as shown in Appendix \ref{appendix:intrinsicSpin}. When the two electrons move away from each other, their motions can be translational such that they have no impact on the orientations of intrinsic angular momenta. Or, their motions can have the same rotational effects on both electrons such that the impacts on the orientations of the intrinsic angular momenta are the same. In either case, the correlation $\theta_B + \theta_A = \pi$ is preserved when the two electrons physically move apart. In the modern Bell test experiment~\cite{Hensen}, the two electrons have already been remotely separated before being prepared to be entangled. The electron entanglement is achieved by entanglement swapping with photons~\cite{Hensen}. Thus, there is no need to worry about the impacts on the correlation $\theta_B + \theta_A = \pi$ due to the movement of electrons.

\added{Second, how is the spin correlation different from a classical correlation? An example of a classical correlation is a pair of shoes placed in two sealed boxes. The two boxes are remotely separated. When an observer opens one box and finds that the shoe is for the left foot, she will immediately predict that the other shoe located remotely is for the right foot. The spin correlation presented here has two important differences from this classical correlation. Firstly, when an observer measures the spin of electron A, the result is completely random. This randomness is intrinsic, while the left or right property of the shoe in the classical correlation example is a deterministic property. The properties correlated in a pair of entangled spins are the spin orientations, and they are random variables. Secondly, to uniquely specify the correlation between two random variables of orientation, one needs to specify the correlation on both the $z$-axis and the $y$-axis, as shown in \eqref{measABz} and \eqref{measABy}.}

\subsection{\added{Time Dependency of Measurement Results on A Pair of Entangled Spins}}
\label{subsec:timeDep}

\added{The spin entanglement described in our model implies an interesting property. In the singlet spin state, even if $\theta_B + \theta_A = \pi$ is preserved before measurement,} if the measurements on $A$ and $B$ are performed at different times, there is a possibility that the entanglement effect can be diminished. Let us next examine this more subtle situation.
 
Suppose that the electron pairs with the correlation $\theta_B + \theta_A=\pi$ established, are separated far away. Electron $A$ is with observer Alice, while electron $B$ is with observer Bob. The correlation $\theta_B + \theta_A=\pi$ is maintained even though the electron pair are space-like separated. At time $t_A^m$, Alice performs measurement on $A$. The probability density is given by \eqref{measA}. Alice will obtain the measurement result randomly. Suppose that she observes spin-down. At this point, from Bob's point of view, if he measures electron $B$ along any direction, he will obtain spin-up or spin-down randomly. Now Alice sends Bob her measurement results along with information on the direction of measurement. With Alice's information, Bob infers that $\theta_A$ trendes downward relative to the measurement direction, and therefore $\theta_B$ trendes upward relative to the same measurement direction due to the correlation $\theta_B + \theta_A=\pi$. Consequently, he predicts that he will obtain spin-up if he performs the measurement on $B$ along the same orientation as Alice. However, due to fluctuation, $\theta_B$ can become trending downward after some time. Bob's measurement must be performed at a time $t_B^m$ before $\theta_B$ becomes tending downward. This constraint can be approximately expressed as
\begin{equation}
    \label{timeRel}
    t_B^m - t_A^m < \tau_B^+,
\end{equation}
where $\tau_B^+$ is the average time $\theta_B$ stays in $\Theta^+_B$ before it switches to $\Theta^-_B$. The scenario is depicted in Figure 6.

Eq. \eqref{timeRel} implies that Bob's prediction on his measurement outcome of spin-up for $B$ is valid only for a period of time. Although the entanglement between the electron pairs is preserved after they are remotely separated, once Alice takes her measurement in $A$, Bob must perform his measurement on $B$ within the time constraint \eqref{timeRel}. This is different from the prediction of standard quantum mechanics, where there is no such time constraint. If $\tau_B^+$ is sufficiently large, our model will practically give the same prediction as standard quantum mechanics.

One may argue that if $t_B^m > t_A^m + \tau_B^+$, Bob will obtain measurement results of spin-down or spin-up, what is the difference between an entangled pair and a non-entangled electron pair? The answer is that for non-entangled electron pairs, there is no correlation between the measurement results of Alice and Bob. Bob cannot make any prediction on the outcome of his measurement of $B$ even immediately after receiving the information from Alice. 

This interesting property of time dependency of the measurement results on a pair of entangled spin will be used to design a modified Bell test in Section \ref{sec:timeDep}.


\begin{figure*}
\begin{center}
\includegraphics[scale=0.75]{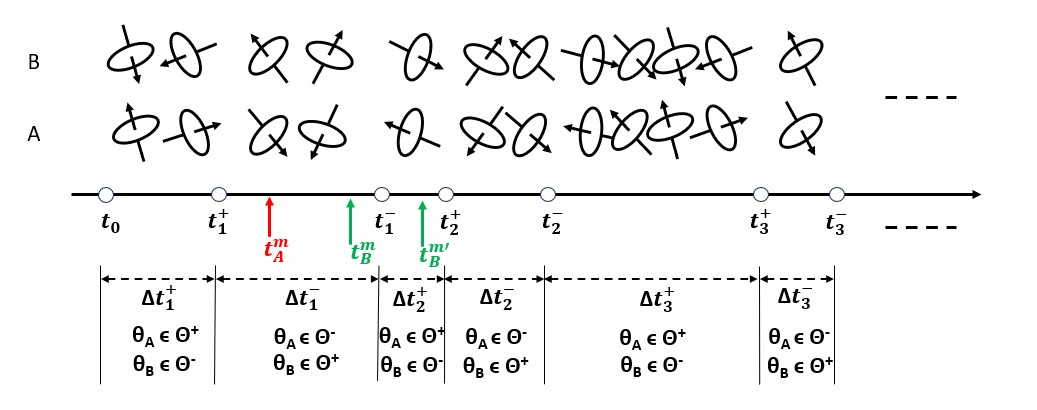}
\caption{Spin entanglement of two electrons from time perspective. Here, the orientation of intrinsic angular momentum for each electron is completely random, but the orientations for both $A$ and $B$ are always opposite to each other. If Alice performs the measurement at time $t_m^A$ that the initial orientation is $\theta_A\in\Theta^-$, the measurement outcome is spin-down. If Bob performs his measurement at $t_m^B < t_1^-$, he will obtain spin-up, as predicted by the singlet spin state. On the other hand, if Bob performs his measurement at $t_m^{B'} > t_1^-$ the measurement outcome can be spin-down or spin-up.}
\end{center}
\label{fig6}
\end{figure*}

\subsection{Other Bell States}
In previous sections, we show that the spin singlet state can be described by \eqref{measABz} and \eqref{measABy}. Here, we would like to recover other Bell states using the probability density functions. If the correlations between the electron pair are such that $\theta_A^z+\theta^z_B=\pi$ along the $z$-axis, and $\theta_A^y=\theta^y_B$ along the $y$-axis, we must also have
\begin{equation}
    \label{measABy2}
    \Tilde{p}(\theta_A^y, \theta_B^y) = \frac{1}{2}\delta(\theta_A^y)\delta(\theta_B^y) + \frac{1}{2}\delta(\theta_A^y-\pi)\delta(\theta_B^y-\pi).
\end{equation}
Thus, measurement along the $y$-axis results in the same spin directions for both electrons, whereas measurement along the $z$-axis yields opposite spin directions. A pair of electrons described by both \eqref{measABz} and \eqref{measABy2} shares the same correlation specified by the spin triplet states in standard quantum mechanics
\begin{equation}
    \label{triplet1}
    |\Psi^+\rangle := \frac{1}{\sqrt{2}}(|\uparrow\rangle_A|\downarrow\rangle_B + |\downarrow\rangle_A|\uparrow\rangle_B).
\end{equation}
Similarly, we can give equivalent descriptions of the other two Bell states in standard quantum mechanics. In the case that the initial orientations of angular momentum of electron $A$ are uniformly distributed for all $\theta\in[0,\pi]$, and the correlation between the electron pair are such that $\theta_A^z=\theta^z_B$ along the $z$-axis, the joint probability density function is
\begin{equation}
    \label{measABz2}
    \Tilde{p}(\theta_A^z, \theta_B^z) = \frac{1}{2}\delta(\theta_A^z)\delta(\theta_B^z) + \frac{1}{2}\delta(\theta_A^z-\pi)\delta(\theta_B^z-\pi). 
\end{equation}
The combination of \eqref{measABz2} and \eqref{measABy} is equivalent to the Bell state 
\begin{equation}
    \label{triplet2}
    |\Phi^-\rangle:=\frac{1}{\sqrt{2}}(|\uparrow\rangle_A|\uparrow\rangle_B - |\downarrow\rangle_A|\downarrow\rangle_B),
\end{equation}
and the combination \eqref{measABz2} and \eqref{measABy2} is equivalent to 
\begin{equation}
    \label{triplet3}
    |\Phi^+\rangle:=\frac{1}{\sqrt{2}}(|\uparrow\rangle_A|\uparrow\rangle_B + |\downarrow\rangle_A|\downarrow\rangle_B).
\end{equation}
We therefore recover the four Bell states.

\subsection{Bell Inequality}
\label{sec:BellIneq}
Bell inequality is developed to prove that local hidden variable theory cannot produce the predictions of standard quantum mechanics. Bell inequality is violated when tested with an entangled pair of photons or electrons, demonstrating that there is non-local correlation in standard quantum mechanics. Although modern Bell experiments~\cite{Hensen} confirm that such a non-local correlation does not imply non-local causality, this Bell non-locality is still daunting to the quantum physics community because there is no intuitive physical model to explain the phenomenon. Here, we investigate what insights our spin model can bring to this challenge.

Consider the experiment performed by Alice and Bob discussed in the previous section using an entangled pair of electrons. The CHSH version of Bell inequality reads
\begin{equation}
    \label{CHSH}
    | E(a, b) - E(a, b') + E(a', b) + E(a', b')| \le 2.
\end{equation}
Here, $a, a'$ are the detector settings for Alice and $b, b'$ are the detector settings for Bob. They usually refer to the directions of the magnetic field in the spin measurement and can be considered as unit vectors for the corresponding directions. The term $E(a, b)$ is the expectation value of the measurement outcomes. That is, the statistical average of $S_A(a)S_B(b)$ where $S_A(a)$ and $S_B(b)$ are the spin measurement results for setting $a$ and $b$, respectively. For the setting $a$, the unit vector for the direction of the magnetic field is $\hat{a}$, which we still denote as $a$ for simplicity of notation. Then, $S_A(a)=+1$ for spin-up and $S_A(b)=-1$ for spin-down. Similar meanings can be inferred for the other $E(\cdot)$ terms in \eqref{CHSH}. 

The expectation value $E(\cdot)$ depends on the state of the entangled pair. Suppose that the entangled electron pair is in the singlet described by the joint probability density \eqref{measABz} and \eqref{measABy}. It is important to note that both \eqref{measABz} and \eqref{measABy} are needed to give a complete description of the singlet state. In Appendix \ref{appendix:correl}, we prove that given the joint probability density \eqref{measABz} and \eqref{measABy},
\begin{equation}
    \label{measCorrel}
    E(a, b) = - (a\cdot b) = -\cos (\gamma),
\end{equation}
where $\gamma$ is the angle between unit vectors $a$ and $b$. This result is exactly the same as the result from standard quantum mechanics. If Alice chooses $a=0$ and $a'=\pi/2$, and Bob chooses $b=\pi/4$ and $b'=3\pi/4$, we will have $| E(a, b) - E(a, b') + E(a', b) + E(a', b')| = 2\sqrt{2} > 2$. Thus, the CHSH inequality is violated.

The reason for the violation of the CHSH inequality is due to the fact that the joint probability density \eqref{measABz} and \eqref{measABy} cannot be factorized. In other words, they cannot be written as a product of two factors, one only contains variable for $A$ and the other only contains variable for $B$. Such a correlation can be maintained even though the two electrons are remotely separated. The root cause of this correlation is that the random variables are correlated, $\theta^z_A+\theta^z_B=\pi$ and $\theta^y_A+\theta^y_B=\pi$. These random variables are intrinsically local to the electrons. The term non-local correlation, or Bell non-locality, is misleading. Instead, as Hall pointed out \cite{Hall2015}, a better term is Bell non-separability.

\begin{figure*}
\begin{center}
\includegraphics[scale=0.75]{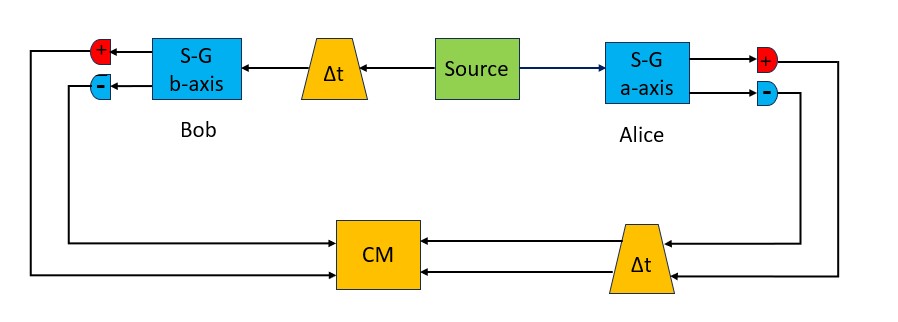}
\caption{Bell test with time delay. Starting with a Bell test experiment setup that has already successfully confirmed the violation of Bell-CHSH inequality. Then, we add a time delay in the measurement. Bob delays his measurement by a time period $\Delta t$, while the signals from Alice's Stern-Gerlach detectors are delayed by the same $\Delta t$ before it is sent to the coincident monitor. When $\Delta t$ is chosen to be sufficiently large, we expect the Bell-CHSH inequality becoming non-violated.}
\end{center}
\label{fig7}
\end{figure*}

\section{Possible Experiments to Confirm the Spin Theory}
\label{sec:PossibleExp}
The spin theory presented here so far recovers many of the results from standard quantum mechanics when certain conditions are taken to the limits. However, when these conditions are not taken to the limits, the present theory will give some results different from those of standard quantum mechanics. In this section, we will provide two possible experiments to confirm the difference.

\subsection{Dependency of Quantization on Interaction with Magnetic Field}
\label{sec:BzDep}
In the derivation of \eqref{pinfty}, one assumes that the parameter $m$, which is related to the order of Tsallis relative entropy as shown in \eqref{alpham}, monotonically increases when the electron travels along an inhomogeneous magnetic field $B_z$. Standard quantum mechanics postulates that quantization of electron spin is an intrinsic property. Practically, it corresponds to the case that whenever an external field $B_z$ is applied, quantization occurs instantaneously, such that only two discrete results can be observed. However, the theory presented here allows $m$ to take a finite number when the magnetic field $B_z$ is applied. In that case, the probability density \eqref{pthetam} should be used instead of the probability density \eqref{pinfty} with two quantized results.

Suppose the weak magnetic field in the Stern-Gerlach apparatus is given by
\begin{equation}
    \label{Bz}
    \Vec{B} =  {B}_0\hat{z} - \eta z\hat{z},
\end{equation}
where ${B}_0\hat{z}$ is a constant field, and the coefficient $\eta$ determines the strength of the gradient of the magnetic field. Since the orientation of the intrinsic magnetic moment of the electron follows a probability density distribution \eqref{pthetam}, the electron passing the Stern-Gerlach apparatus will experience a force that also depends on the orientation. Between angles $\theta$ and $\theta+d\theta$, the force is
\begin{equation}
    \label{force}
    \Vec{F}(\theta) = -\nabla (p_m(\theta)\Vec{\mu}\cdot \Vec{B}) = \frac{e\hbar\eta}{2m_eZ_m}\cos^{2m+1}(\theta)\hat{z},
\end{equation}
where $m_e$ is the mass of electrons. After traveling in the Stern-Gerlach apparatus for a period of time $\Delta T$, the electron exhibits a displacement along the $z$-axis given by
\begin{equation}
    \label{zdisplacment}
    \Delta z = \frac{e\hbar\eta}{4m_e^2Z_m}(\Delta T)^2 \cos^{2m+1}(\theta).
\end{equation}
The above calculation ignores the effect of the translational component of vacuum fluctuations and only computes the displacement along the $z$-axis using classical mechanics. In theory, this approximation is reasonable when $\Delta T$ is sufficiently large so that $\Delta z$ is much larger than the effect of the translational component of vacuum fluctuations. Eq. \eqref{zdisplacment} shows that the electrons passing through the Stern-Gerlach apparatus will reach the detector screen with a continuous distribution, as shown in Fig. 2b. 

However, as $\Delta T$ increases, the parameter $m$ also increases monotonically, the measurement results may rapidly converge to two discrete lines and yield the same prediction as standard quantum mechanics. Such an experiment can be challenging to implement. The next proposed experiment can be more realistic. 


\subsection{Bell Test with Time Delay}
\label{sec:timeDep}
In Section \ref{sec:SpinEnt}, we show that for an entangled electron pair described by the joint probability density \eqref{measABz} and \eqref{measABy}, the spin correlation can be preserved without a time constraint. However, as explained in Section \ref{subsec:timeDep}, when Alice performs her measurement on the spin of electron $A$, obtains spin-up, and makes her prediction on Bob's measurement result of spin-up along the same direction, the prediction is only valid for a limited time. That is, there is a time constraint specified in \eqref{timeRel} that Bob needs to perform his measurement on the spin of the electron $B$ to confirm Alice's prediction. If $t_m^B > t_m^A + \Delta t^B_+$, Bob will instead obtain the spin-down measurement result. A singlet spin state is described by two probability density functions \eqref{measABz} and \eqref{measABy}. After Alice performs her measurement on the electron $A$ and obtains the result of spin-up, if Bob delays his measurement by a time such that the correlation described by either \eqref{measABz} or \eqref{measABy} is no longer satisfied, the singlet state is degraded and no longer valid. 

We can explore the effect of entanglement degradation for a singlet state due to Bob's delayed measurement in the Bell test experiment, as illustrated in Figure 7. In a Bell test experiment, the electron source generates $N$ copies of pairs of entangled electrons in a singlet state. For each pair, electrons $A$ and $B$ are sent to two Stern-Gerlach apparatuses that are remotely separated\footnote{As mentioned earlier, in modern Bell test experiment~\cite{Hensen}, the two electrons are already separated remotely before being prepared to be entangled. The electron entanglement is achieved by entanglement swapping with photons. There is no need to move the electrons and, therefore, no need to worry about the impacts on the correlation resulting from the movement of the electrons.}. Alice measures the spin of an electron $A$ along an axis labeled by unit vector $a$. At the same time, Bob measures the spin of the electron $B$ along an axis labeled by unit vector $b$. The detectors in each Stern-Gerlach apparatus detect the spin-up and spin-down results and generate the corresponding signals that are sent to the coincidence monitor. Counting the four types $(++, +-, -+, --)$ of coincidences for the $N$ copies of the singlet spin pairs, one can calculate whether the Bell-CHSH inequality is violated. As shown in Section \ref{sec:BellIneq}, for the singlet spin pair described by \eqref{measABz} or \eqref{measABy}, if Alice chooses $a=0$ and $a'=\pi/2$, and Bob chooses $b=\pi/4$ and $b'=3\pi/4$, the maximum violation of the Bell-CHSH inequality is achieved.

Using exactly the same experiment setup, we add an extra step here. Instead that Bob performs the measurement on electron $B$ at the same time as Alice's measurement, he delays his measurement by a time period of $\Delta t$. When $\Delta t$ is sufficiently large, the entanglement correlation of a singlet state is degraded, and we expect that the Bell-CHSH inequality will not be violated. Since we do not know how long to delay in order to observe the non-violation of Bell-CHSH inequality, we can start with a very small delay $\Delta t_1$ such that the Bell-CHSH inequality is still violated, and then gradually increase the delay. We repeat the experiments $M$ times, but choose delay times such that $\Delta t_1 < \Delta t_2 < \ldots < \Delta t_M$. Each experiment consumes $N$ copies of singlet electron pairs. According to the present theory, we expect that the violation of Bell-CHSH inequality will be gradually reduced, and at some time the results will satisfy the Bell-CHSH inequality. The reason for this is that with a sufficiently large delay $\Delta t$, the orientation of the intrinsic angular momentum of the electron $B$ can fluctuate to switch from trending upward to trending downward, or vice versa, as depicted in Figure 4. When an orientation switching occurs, either $\theta_B\in\Theta_B^+\to\theta_B\in\Theta_B^-$, or $\theta_B\in\Theta_B^-\to\theta_B\in\Theta_B^+$, the initial correlation between $\theta_A$ and $\theta_B$ is changed. As $\Delta t$ increases, the orientation of the electron $B$ appears again as a random variable for the $N$ copies of the pairs of electrons, so the initial correlation between $\theta_A$ and $\theta_B$ no longer holds. Consequently, the Bell-CHSH inequality becomes satisfied. Appendix \ref{appendix:correl} gives the mathematical calculation of such a scenario.

The reason why the entanglement relation between $A$ and $B$ is degraded is due to the orientation fluctuations of the intrinsic angular momentum. This is different from the decoherence theory, in which the entanglement relationship between $A$ and $B$ can be destroyed by environmental disturbance. Here, however, the degradation of entanglement during measurement is intrinsic to the electron pair, nothing to do with the environment. Without the measurement event, the entanglement relation between $A$ and $B$ can be preserved without a time limit. Certainly, if there is also environmental disturbance, the entanglement relationship will be destroyed without measurement. So, the present theory is not incompatible with the decoherence theory. However, if confirmed by experiment, the intrinsic degradation of entanglement during measurement can have a non-negligible implication on its applications in quantum computing and quantum information. 

The challenges in performing the above experiment come from choosing the appropriate time delay $\Delta t$. To meet the locality condition of the Bell test, $\Delta t$ should be less than the time it takes for a light signal to travel from Alice to Bob. This is to eliminate the potential loophole of hidden signals sent from Alice to Bob. But $\Delta t$ can also not be too small so that the orientation of intrinsic angular momentum of $B$ switches from $\theta_B\in\Theta_B^+\to\theta_B\in\Theta_B^-$, or $\theta_B\in\Theta_B^-\to\theta_B\in\Theta_B^+$. To resolve this contention, one can separate electrons $A$ and $B$ with a very long distance so that the time window to comply with the locality condition is large enough. Denote the time for a light signal to travel between Alice and Bob as $\tau^{AB}$, then the time delay for Bob to perform his measurement must meet the following constraint,
\begin{equation}
    \tau^B_+ < \Delta t < \tau^{AB}.
\end{equation}


\section{Discussion and conclusions} 
\label{sec:discussion}

\subsection{Conceptual Implications}
In this paper, an intuitive physical model is provided to understand the phenomenon of entanglement between two spins of electrons. As shown in Section \ref{sec:entanglement}, entanglement is simply the manifestation of the correlation between the orientations of intrinsic angular momenta of two electrons. Even though the orientations are random variables, they can have a correlation because of previous interactions between them or because of previous interactions with common external fields. Once the correlation of orientations between two electrons is established, it can be maintained even if the two electrons are remotely separated, until a measurement is performed on them. Performing a measurement on one electron will not cause any change on the other electron. However, the measurement outcome reveals the orientation of the measured electron's intrinsic angular momentum, which enables us to infer the spin measurement outcome of the other electron because of the correlation on the orientations. The seemingly non-local correlation has nothing to do with any non-local causal relation. Although classical correlation can also demonstrate non-local and non-causal relationship, as discussed in Section \ref{sec:SpinEnt}, the fundamental difference in spin correlation is that the correlated variables (the spin orientations) are intrinsically random.

To further clarify the point that there is no non-causal effect in the Bell experiment with a pair of electrons in the singlet state, suppose after Alice performs her experiment and obtains the result of spin-up for electron $A$, she does not send the measurement outcome to Bob. Since the measurement action of Alice is local and has no influence on the electron $B$, from Bob's point of view, he still perceives both $A$ and $B$ in the original singlet state described by \eqref{measABz} and \eqref{measABy}. Alice's and Bob's descriptions on $B$ are different, but both are valid. This is consistent with the relational quantum mechanics (RQM) interpretation advocated by Rovelli~\cite{Rovelli:1995fv, Rovelli07}. Since Bob does not know the measurement outcome from Alice, he predicts that if he measures the electron $B$ along any direction, he will obtain spin-up and spin-down with equal probability. However, if Alice sends the measurement results to Bob, Bob can infer the orientation of the intrinsic angular momentum of electron $B$. Bob's knowledge about electrons $B$ is updated with the new information from Alice. Thus, he can predict that if he measures electron $B$ along the same direction as Alice's measurement, he will obtain the result of spin-down. This analysis also illustrates that the wave function in quantum mechanics is just a mathematical tool and is not associated with certain ontic properties. 

We see that the spin model and theory presented here give an intuitive physical picture that are consistent with several quantum mechanics interpretations. In addition, our theory supports the idea that the measurement outcome is context dependent. The measurement outcome for the spin of an electron depends on the measurement context, that is, the direction of magnetic field used in the Stern-Gerlach apparatus. The two-level discreteness of the measurement outcome is the result of a sufficiently strong gradient of magnetic field. For an electron with initially completely random orientation of intrinsic angular momentum, we can measure the spin along any direction, but always get spin-up and spin-down with equal probability. There is no predefined spin property without specifying an experimental context.

\subsection{Limitations}
\label{subsec:limitations}
There are several limitations that we need to point out for future investigations. The first challenge is to explain why the magnitude of intrinsic angular momentum of an electron is $\hbar/2$. Although Appendix \ref{appendix:intrinsicSpin} gives a reasonable derivation, which is also based on the extended least action principle, there is a strong assumption on random circular motion due to vacuum fluctuation. A more intuitive model with weaker assumptions is desirable.

Secondly, Assumption 4 suggests that when the interaction between the electron and an inhomogeneous magnetic field takes place in a Stern-Gerlach apparatus, the order of relative entropy monotonically increases, so that the direction of intrinsic angular momentum tends to align with the direction of the magnetic field. This assumption seems quite arbitrary. Here, we give a possible intuitive explanation. The key is not to consider the electron as an idealized point particle. Instead, it has non-zero size and is distributed with a spatial volume. We do not assume that it is a rigid rotating body. Instead, it can be divided into many small mass elements $\delta m$. Each mass element is also a charge element $\delta e$, and experiences Larmor precession. The Larmor precession frequency is $\omega = \delta e B_z/\delta m$. In a homogeneous magnetic field, $\omega$ is a constant for each charge element and independent of angle $\theta$. In this case, the probability density function $p_m(\theta)$ should remain unchanged for the entire electron. But in an inhomogeneous field $B_z$, because the electron has spatial size, each charge element will experience a different magnitude of the magnetic field. Consequently, different parts of the electron will have different precession frequencies with $\Delta\omega\propto\Delta B_z$. The upper parts of the electron along the direction of $B_z$ tend to precess faster than the lower parts of the electron. Effectively, the orientation of the overall intrinsic angular momentum is ``pulled" closer to align with the $z$-axis. This corresponds to choosing a larger value of $m$. Spin theory based on stochastic mechanics~\cite{Beyer} seems to give similar results. 

To estimate $\Delta\omega$, consider that in a typical Stern-Gerlach experiment, the gradient of magnetic field is of the order of $10^3$ Tesla/m. The size of the electron is of the order of $10^{-18}$ m. Thus, the difference in magnetic field experienced at the upper part and the lower part of the electron is about $\Delta B_z=10^{-15}$ Tesla. Then, the Larmor precession frequency difference $\Delta\omega$ is on the order of $10^{-3}$Hz. On the other hand, the typical Larmor frequency $\omega$ for an electron is $10^{10}$Hz. Thus, $\Delta\omega/\omega = 10^{-13}$. This is an extremely small ratio. It raises the question whether such a small ratio can cause the orientation of the overall intrinsic angular momentum to tilt closer to align with the $z$-axis. A more rigorous theory is needed to confirm this assumption. Either considering the electron as an idealized charged point particle or as a charge body with finite spatial size, seems insufficient as an accurate physical model for the electron. We speculate that using field theory could be a better framework to describe the physical model of an electron, which is beyond the scope of this work but has been studied in \cite{Sebens}.


Lastly, the Schr\"{o}dinger-Pauli equation derived in Section \ref{sec:PauliEq} is only valid when the external magnetic field is along the $z$-axis. We need to extend the derivation for an external magnetic field with arbitrary direction. 

\subsection{Conclusions}
From the extended least action principle, and with additional assumptions on the intrinsic angular momentum of an electron, particularly its random orientations, we are able to develop a theory that recovers the quantum properties for electron spin. The key component of the theoretical framework here is the introduction of relative entropy for both translational and rotational random fluctuations. The two-level quantization of spin measurement results is a mathematical consequence when the order of relative entropy approaches a limit. We also obtain the same formulation as standard quantum mechanics when the direction of magnetic field of a second Stern-Gerlach apparatus is rotated with an angle from the direction of magnetic field of the first Stern-Gerlach apparatus. Recursively applying the extended least action principle, we have derived the Schr\"{o}dinger-Pauli equation. 

An important result here is that we provide an intuitive physical model and formulation to explain the entanglement phenomenon between two electron spins. The root cause of entanglement is the correlation between the orientations of the intrinsic angular momenta of the two electrons. The correlation is established between two random variables due to previous interactions. Since the orientation of the angular momentum is an intrinsic local property of the electron, the correlation can be maintained even though the two electrons are space-like separated and have no further interaction. Mathematically, we give an equivalent formulation of the four Bell states using probability density function instead of wave function. Using the probability density functions for a pair of entangled electrons, we prove that the Bell-CHSH inequality is violated. The violation of Bell-CHSH inequality is due to the same root cause of entanglement, that is, correlation of the orientations of the angular momenta which is preserved even when the two electrons are separated.

When certain conditions are not taken to the limits, the present theory gives certain results that are different from standard quantum mechanics. For instance, if the Stern-Gerlach apparatus is set up with a sufficiently weak gradient of the magnetic field along the $z$-axis, we expect the electron detector screen to exhibit a continuous distribution of displacements along the $z$-axis, instead of only two discrete lines. Another more interesting experiment proposed is to add a delay before Bob's measurement in the typical Bell test experiment, which could result in the non-violation of the Bell-CHSH inequality. The second experiment shows the effect of time dependency of the measurement results involving spin entanglement. If confirmed by experiment, such an effect can impose a limitation on the application of spin entanglement in quantum computing and quantum information.

The interplay of the extended least action principle and the information metrics for vacuum fluctuations is proved to be a valuable theoretical framework. We have shown that the theoretical framework based on the extended least action principle has successfully recovered non-relativistic quantum mechanics~\cite{Yang2023}, the relativistic quantum scalar field theory~\cite{Yang2024}, and in this paper the quantum theory for electron spin. We believe that the results in this paper are interesting because we can explain the entanglement phenomenon with an intuitive physical model, and the spin-entanglement model implies potential new results that are falsifiable with the modified Bell test with time delay. 

There are clear limitations in the present theory with the assumptions on the detailed physical model of electron spin. A more rigorous theory must be developed to further justify these assumptions. This paper is intended to be an intermediate step in the investigation of spin theory using information metrics of vacuum fluctuations.

\begin{acknowledgements}
\added{
The author thanks the anonymous referees for their valuable comments, which motivate the author to clarify the choice of the Tsallis divergence and the difference between classical correlation and spin entanglement.}
\end{acknowledgements}




\onecolumngrid

\pagebreak

\appendix

\section{Deriving Basic Quantum Formulation From the Extended Least Action Principle}
\label{sec:shorttime}
Basic quantum formulation can be derived by recursively applying the extended least action principle in two steps. First, we consider the dynamics of a system with an infinitesimal time internal $\Delta t$ due to only vacuum fluctuation. Define the probability that the system will transition from a 3-dimensional space position $\mathbf{x}$ to another position $\mathbf{x}+\mathbf{w}$, where $\mathbf{w}=\Delta \mathbf{x}$ is the displacement in a 3-dimensional space due to fluctuations, as $\wp(\mathbf{x}+\mathbf{w}|\mathbf{x})d^3\mathbf{w}$. The expectation value of the classical action is $S_c=\int \wp(\mathbf{x}+\mathbf{w}|\mathbf{x})Ld^3\mathbf{w}dt$. Since we only consider the vacuum fluctuations, the Lagrangian $L$ only contains the kinetic energy, $L=\frac{1}{2}m\mathbf{v}\cdot\mathbf{v}$. For an infinitesimal time internal $\Delta t$, one can approximate the velocity $\mathbf{v}=\mathbf{w}/\Delta t$. This gives 
\begin{equation}
\label{action1}
    S_c=\frac{m}{2\Delta t}\int^{+\infty}_{-\infty} \wp(\mathbf{x}+\mathbf{w}|\mathbf{x})\mathbf{w}\cdot\mathbf{w}d^3\mathbf{w}.
\end{equation}
The information metrics $I_f$ is defined as the Kullback–Leibler divergence, to measure the information distance between $\wp(\mathbf{x}+\mathbf{w}|\mathbf{x})$ and a uniform prior probability distribution $\mu$ that reflects the vacuum fluctuations are completely random with maximal ignorance~\cite{Caticha2019, Jaynes}, 
\begin{align*}
    I_f  &=: D_{KL}(\wp(\mathbf{x}+\mathbf{w}|\mathbf{x}) || \mu) \\
    &= \int \wp(\mathbf{x}+\mathbf{w}|\mathbf{x})ln[\wp(\mathbf{x}+\mathbf{w}|\mathbf{x})/\mu]d^3\mathbf{w}.
\end{align*}
Insert both $S_c$ and $I_f$ into (\ref{totalAction}) and perform the variation procedure, one obtain
\begin{equation}
\label{transP}
    \wp(\mathbf{x}+\mathbf{w}|\mathbf{x}) = \frac{1}{Z}e^{-\frac{m}{\hbar\Delta t}\mathbf{w}\cdot\mathbf{w}},
\end{equation}
where $Z$ is a normalization factor. Equation (\ref{transP}) shows that the transition probability density is a Gaussian distribution. The variance for the vector component $w_i$ is $\langle w_i^2\rangle = \hbar\Delta t/2m$, where $i\in\{1, 2, 3\}$ denotes the spatial index. Recalling that $w_i/\Delta t = v_i$ is the approximation of velocity due to the vacuum fluctuations, one can deduce
\begin{equation}
\label{exactUR}
    \langle\Delta x_i\Delta p_i\rangle = \frac{\hbar}{2}.
\end{equation}
Applying the Cauchy–Schwarz inequality gives
\begin{equation}
    \langle\Delta x_i\rangle\langle\Delta p_i\rangle \ge \hbar/2.
\end{equation}

In the second step, we will derive the dynamics for a cumulative period from $t_A\to t_B$. In classical mechanics, the equation of motion is described by the Hamilton-Jacobi equation, 
\begin{equation}
    \label{HJE}
    \frac{\partial S}{\partial t }+ \frac{1}{2m}\nabla S\cdot\nabla S + V = 0.
\end{equation}
Suppose that the initial condition is unknown and define $\rho ({x}, t)$ as the probability density for finding a particle in a given volume of the configuration space. The probability density must satisfy the normalization condition $\int \rho (\mathbf{x}, t) d^3\mathbf{x} = 1$, and the continuity equation 
\begin{equation*}
    \frac{\partial\rho (\mathbf{x}, t)}{\partial t }+ \frac{1}{m}\nabla \cdot(\rho (\mathbf{x}, t)\nabla S) = 0.
\end{equation*}
The pair $(S, \rho)$ completely determines the motion of the classical ensemble. As pointed out by Hall and Reginatto~\cite{Hall:2001,Hall:2002}, the Hamilton-Jacobi equation, and the continuity equation, can be derived from classical action
\begin{equation}
    \label{cAction}
    A_c = \int\rho\{ \frac{\partial S}{\partial t} + \frac{1}{2m}\nabla S\cdot\nabla S + V\} d^3\mathbf{x}dt
\end{equation}
through fixed point variation with respect to $\rho$ and $S$, respectively. Note that $A_c$ and $S$ are different physical variables, where $A_c$ can be considered as the ensemble average of classical action and $S$ is a generation function that satisfies $\mathbf{p}=\nabla S$~\cite{Yang2023}. 

To define the information metrics for the vacuum fluctuations, $I_f$, we slice the time duration $t_A\to t_B$ into $N$ short time steps $t_0=t_A, \ldots, t_j, \ldots, t_{N-1}=t_B$, and each step is an infinitesimal period $\Delta t$. In an infinitesimal time period at time $t_j$, the particle not only moves according to the Hamilton-Jacobi equation but also experiences random fluctuations. Such additional revelation of distinguishability due to the vacuum fluctuations on top of the classical trajectory is measured by the following definition,
\begin{align}
\label{DLDivergence}
    I_f &=: \sum_{j=0}^{N-1}\langle D_{KL}(\rho (\mathbf{x}, t_j) || \rho (\mathbf{x}+\mathbf{w}, t_j))\rangle_w \\
    &=\sum_{j=0}^{N-1}\int d^3\mathbf{w} d^3 \mathbf{x}\wp(\mathbf{x}+\mathbf{w}| \mathbf{x})\rho (\mathbf{x}, t_j)ln \frac{\rho (\mathbf{x}, t_j)}{\rho (\mathbf{x}+\mathbf{w}, t_j)}.
\end{align}
When $\Delta t\to 0$, $I_f$ turns out to be~\cite{Yang2023}
\begin{equation}
\label{FisherInfo}
    I_f = \int d^3\mathbf{x}dt \frac{\hbar}{4m}\frac{1}{\rho}\nabla\rho \cdot \nabla\rho.
\end{equation}
Eq. (\ref{FisherInfo}) contains the term related to Fisher information for the probability density~\cite{Frieden} but is much more physical significant than Fisher information. 
Inserting (\ref{cAction}) and (\ref{FisherInfo}) into (\ref{totalAction}), and performing the variation procedure on $I$ with respect to $S$ gives the continuity equation, while variation with respect to $\rho$ leads to the quantum Hamilton-Jacobi equation,
\begin{equation}
\label{QHJ}
    \frac{\partial S}{\partial t} + \frac{1}{2m}\nabla S\cdot\nabla S + V - \frac{\hbar^2}{2m}\frac{\nabla^2\sqrt{\rho}}{\sqrt{\rho}} = 0,
\end{equation}
Defined a complex function $\Psi=\sqrt{\rho}e^{iS/\hbar}$, the continuity equation and the extended Hamilton-Jacobi equation (\ref{QHJ}) can be combined into a single differential equation,
\begin{equation}
    \label{SE}
    i\hbar\frac{\partial\Psi}{\partial t} = [-\frac{\hbar^2}{2m}\nabla^2 + V]\Psi,
\end{equation}
which is the Schr\"{o}dinger Equation. 

The last term in (\ref{QHJ}) is the Bohm quantum potential~\cite{Bohm1952}. The Bohm potential is considered responsible for the non-locality phenomenon in quantum mechanics~\cite{Bohm2}. Historically, its origin is mysterious. Here we show that it originates from information metrics related to relative entropy, $I_f$. 

As noted in \cite{Yang2023}, the choice of relative entropy for $I_f$ can be Renyi divergence or Tsallis divergence, which results in a family of Schr\"{o}dinger equations that depends on the order of relative entropy $\alpha$. When $\alpha=1$, the regular Schr\"{o}dinger equation is recovered. The flexibility to choose general relative entropy definitions is a very useful mathematical tool for further theoretical investigation of more advanced quantum theory, as we have shown in the present work.

\section{Intrinsic Angular Momentum Due to Vacuum Fluctuations}
\label{appendix:intrinsicSpin}
Assumption 1 on vacuum fluctuations in Section \ref{LIP} does not specify the details of the fluctuating motion. The electron exhibits random displacement due to vacuum fluctuation, and such displacement is described as a vector. The displacement vector is intrinsically local in the sense that it is independent of the reference of origin of the global coordinate system. Given a local point and an infinitesimal time period, the displacement vector can be just a translational motion, i.e. the direction of velocity is in parallel with the displacement vector. However, there is no reason that the direction of velocity must be in parallel with the displacement vector. When the velocity vector has a component perpendicular to the displacement vector, the electron will exhibit rotational movement that can be characterized by an angular momentum. Here we will further assume that rotational movement is circular. Thus, the vacuum fluctuation not only causes the translational displacement, but also comprises components of motion such as circular rotations. Essentially, the vacuum fluctuation produces virtual circular rotations with random radius. This model has been proposed in \cite{Niehaus16, Niehaus22}, but no derivation of the magnitude of the intrinsic angular momentum has been provided. Here, with the help of the extended least action principle, we will show that such random circular rotations give rise to the intrinsic angular momentum of magnitude $\hbar/2$. 

Denote the radius vector of the circular motion in an infinitesimal time period $\Delta t$ as $\Vec{u}$ and the velocity vector as $\dot{\Vec{u}}$. Here we will only consider the circular motion of the electron such that $\dot{\Vec{u}}$ is perpendicular to $\Vec{u}$. The angular momentum is the cross product of $\Vec{u}$ and $m\dot{\Vec{u}}$
\begin{equation}
   \vec{L}_s = \Vec{u} \times m\dot{\Vec{u}}.
\end{equation}
The orientation of $\vec{L}_s$ is completely random. We choose the coordinate plane perpendicular to $\vec{L}_s$ for a further detailed analysis of the circular motion. Here, the radius $u$ is a random variable. Let $\omega$ be the angular frequency. Then, the velocity magnitude is $\dot{u} = \omega u$, and the angular momentum magnitude is $L_s = m\omega u^2$. Since the radius $u$ is a random variable, we denote $p(u)$ as the probability density of the radius $u$. It must satisfy the normalization condition 
\begin{equation}
    \int_0^{\infty} p(u)du  = 1.
\end{equation}
For this circular motion, the Lagrangian $L$ contains only the kinetic energy $L=\frac{1}{2}m(\omega u)^2$. For an infinitesimal period of time $\Delta t$, the expectation value of classical action is 
\begin{equation}
\label{A3}
    A_c = \frac{m}{2}\int_0^{\infty}\int_0^{\Delta t} p(u)(\omega u)^2 du dt.
\end{equation}
Note that $d\varphi = \omega dt$ where $\varphi$ is the angle of circular rotation. In an infinitesimal period of time $\Delta t$, the electron rotates $\Delta\varphi = \omega\Delta t$. We can rewrite (\ref{A3}) as
\begin{equation}
\label{A4}
    A_c = \frac{m}{2}\int_0^{\infty}\int_0^{\Delta \varphi} p(u)\omega u^2 du d\varphi.
\end{equation}
Define the information metrics exhibited during this time period as a relative entropy
\begin{equation}
    I_f := \int_0^{\infty}\int_0^{\Delta \varphi} p(u)ln\frac{p(u)}{\mu} dud\varphi,
\end{equation}
where $\mu$ is a uniform probability density to reflect the total ignorance of knowledge due to complete randomness of fluctuations. Then, the total action, per (\ref{totalAction}), is
\begin{equation}
    A_t = A_c + \frac{\hbar}{2}I_f = \frac{m}{2}\int p(u)\omega u^2 du d\varphi + \frac{\hbar}{2} \int p(u)ln\frac{p(u)}{\mu} dud\varphi.
\end{equation}
Taking the variation of $A_t$ over the functional variable $p(u)$, and demanding $\delta A_t = 0$, we obtain
\begin{equation}
    \frac{m}{2}\omega u^2 + \frac{\hbar}{2}(ln\frac{p(u)}{\mu} + 1) =0.
\end{equation}
This gives
\begin{equation}
    p(u) = \frac{1}{Z}e^{-\frac{m\omega}{\hbar}u^2},
\end{equation}
where $Z$ is the normalization factor. We can then compute the variance of $u$ as
\begin{equation}
    \langle u^2\rangle = \int p(u)u^2 du = \frac{\hbar}{2m\omega}.
\end{equation}
This gives the expectation value of local angular momentum
\begin{equation}
    \langle L_s\rangle = \langle m\omega u^2\rangle = \frac{\hbar}{2}.
\end{equation}
Thus, by assuming that vacuum fluctuations cause random circular motions for the electron, we show that the electron possesses an intrinsic angular momentum with magnitude of $\hbar / 2$. Such intrinsic angular momentum is local in the sense that it is independent of the global orbital movement. Its orientation is completely random. These properties have been summarized as Assumption 3 in Section \ref{sec:SpinTheory}. But based on the derivation showing in this appendix, we can replace Assumption 3 with
\begin{displayquote}
\emph{Assumption 3a -- The vacuum fluctuations cause an electron in free space to exhibit random circular movements in addition to random translational movements.}
\end{displayquote}
Since the assumption of circular movements appears to be a very strong assumption, it is more prudent to just adopt Assumption 3. We speculate that classical field theory could be a better framework to describe these random movements more accurately, which is beyond the scope of this work but has been studied in \cite{Sebens}.

\section{\added{The Choice of Relative Entropy}}
\label{sec:choosingI_f}
\added{
Eq. (\ref{I_f}) is defined with the Tsallis divergence, a one-parameter generalization of the Kullback-Leilber divergence. Alternatively, we can define $I_f$ with another popular one-parameter generalization of the Kullback-Leilber divergence, the R\'{e}nyi divergence. Here we show that defining $I_f$ with the R\'{e}nyi divergence leads to the same results in spin quantization. The R\'{e}nyi divergence is defined as
\begin{equation}
\label{I_f2}
    I^R_f = \frac{1}{\alpha-1}ln(\int_0^{\pi}\int_0^{\Delta \varphi} \frac{p^{\alpha}(\theta)}{\sigma^{\alpha-1}}d\theta d\varphi).
\end{equation}
The total action in \eqref{totalActionT} becomes
\begin{equation}
\label{totalActionR}
    A_t = -\frac{1}{2}g_sL_s\int p(\theta) \cos(\theta)d\theta d\varphi  + \frac{\hbar}{2(\alpha-1)}ln(\int\frac{p^{\alpha}(\theta)}{\sigma^{\alpha-1}}d\theta d\varphi).
\end{equation}
Taking the variation of $A_t$ over the functional variable $p(\theta)$, and demanding $\delta A_t = 0$, we obtain
\begin{equation}
\label{C4}
    -\frac{1}{2}g_sL_s\cos(\theta) + \frac{\alpha\hbar}{2(\alpha-1)F_{\alpha}[p]}[\frac{p(\theta)}{\sigma}]^{\alpha -1} =0,
\end{equation}
where $F_{\alpha}[p]$ is a functional that depends on the functional form of $p$ but not on the parameter $\theta$ and $\varphi$,
\begin{equation}
    F_{\alpha}[p] = \int\frac{p^{\alpha}(\theta)}{\sigma^{\alpha-1}}d\theta d\varphi.
\end{equation}
Given the normalization condition, one recognize that
\begin{equation}
    \lim_{\alpha\to 1}F_{\alpha}[p] = \int p\theta d\varphi = 1,
\end{equation}
which is independent of the functional form of $p$.
From \eqref{C4}, we obtain
\begin{equation}
    \label{ptheta2}
    p(\theta) = \sigma \{\frac{(\alpha-1)g_sL_s}{\alpha\hbar}F_{\alpha}[p]\cos(\theta)\}^{\frac{1}{\alpha -1}} =\frac{1}{Z_{\alpha}}\{F_{\alpha}[p]\cos(\theta)\}^{\frac{1}{\alpha -1}},
\end{equation}
Using the definition \eqref{alpham}, the probability density \eqref{ptheta2} is rewritten as
\begin{equation}
    \label{pthetam2}
    p_m(\theta)=\frac{1}{Z_m}(F_m[p]\cos(\theta))^{2m}, m \in \mathbb{N}.
\end{equation}
Since $\lim_{m\to\infty}F_m[p]=1$, which is independent of the functional form of $p$, we still have
\begin{equation}
        \lim_{m\to\infty}p_m(\theta)  \propto \left\{ \begin{array}{rcl}  0, & \mbox{for} & \theta\in(0, \pi) \\
         1, & \mbox{for} & \theta = 0, \pi \end{array}\right.
\end{equation}
which is the same as \eqref{pinfty}. It shows that only two discrete measurement results can be obtained.}

\added{To conclude the discussion on the choice of relative entropy, suppose that we choose the K-L divergence, the total action becomes
\begin{equation}
\label{totalActionR}
    A_t = -\frac{1}{2}g_sL_s\int p(\theta) \cos(\theta)d\theta d\varphi  + \frac{\hbar}{2}\int p(\theta)ln\frac{p(\theta)}{\sigma}d\theta d\varphi.
\end{equation}
Performing the variation procedure, we eventually obtain
\begin{equation}
    p(\theta) = \frac{1}{Z}e^{\frac{g_sL_s}{\hbar}\cos(\theta)}.
\end{equation}
This form of $p(\theta)$ does not lead to the conclusion of discrete measurement results. Although the K-L divergence is the limit of the Tsallis divergence or the R\'{e}nyi divergence when $m\to\infty$ (or $\alpha\to 1$), note that in obtaining \eqref{pinfty}, we first perform the variation procedure to obtain $p_m(\theta)$, then take the limit $m\to\infty$. Switching the order of these two mathematical steps does not lead to the same result. Therefore, K-L divergence is not an appropriate choice.}

\section{Proof of \eqref{pupbeta2} in Standard Quantum Mechanics}
\label{appendix:rotation}
In standard quantum mechanics, if an electron is measured with result of spin-up along a direction that is tilt an angle $\beta_1$ from the $z$-axis, its state can be described as
\begin{equation}
    \label{rotationUp1}
    |\psi_{\beta_1}^+\rangle = \cos(\beta_1/2)|\uparrow\rangle + \sin(\beta_1/2)|\downarrow\rangle.
\end{equation}
This state can be obtained by rotating the state of spin-up along the $z$-axis by an angle $\beta_1$. The rotation operator is $\hat{R}(\beta_1) = e^{-i\beta_1\sigma_z/2}$ where $\sigma_z$ is the $z$ component of the Pauli operator. Suppose the electron is in this initial state, and passes a Stern-Gerlach apparatus that is configured with a magnetic field with direction tilt an angle $\beta_2$ from the $z$-axis. The measurement operator for spin-up along the $\beta_2$ direction is $\hat{O}_{+} = |\psi_{\beta_2}^+\rangle\langle\psi_{\beta_2}^+|$ where
\begin{equation}
    \label{rotationUp1}
    |\psi_{\beta_2}^+\rangle = \cos(\beta_2/2)|\uparrow\rangle + \sin(\beta_2/2)|\downarrow\rangle.
\end{equation}
The probability is then given by
\begin{equation}
    \label{probUp_beta12}
    p(\uparrow|\beta_2, \beta_1) = |\langle\psi_{\beta_1}^+|\psi_{\beta_2}^+\rangle|^2 = \cos^2((\beta_2-\beta_1)/2).
\end{equation}
The measurement operator for spin-down along the $\beta_2$ direction is $\hat{O}_{-} = |\psi_{\beta_2}^-\rangle\langle\psi_{\beta_2}^-|$ where
\begin{equation}
    \label{rotationUp1}
    |\psi_{\beta_2}^-\rangle = \sin(\beta_2/2)|\uparrow\rangle - \cos(\beta_2/2)|\downarrow\rangle.
\end{equation}
The probability is then given by
\begin{equation}
    \label{probDown_beta12}
    p(\downarrow|\beta_2, \beta_1) = |\langle\psi_{\beta_1}^+|\psi_{\beta_2}^-\rangle|^2 = \sin^2((\beta_2-\beta_1)/2).
\end{equation}
Eqs. \eqref{probUp_beta12} and \eqref{probDown_beta12} are the same as \eqref{pupbeta2}.

\section{Proof of \eqref{measCorrel}}
\label{appendix:correl}
For a spin singlet state, the correlation can be described as $\theta_A^z+\theta_B^z=\pi$ and $\theta_A^y+\theta_B^y=\pi$. The joint probability density functions are given by \eqref{measABz} and \eqref{measABy}. The physical implication is that if measurement of $A$ along the $z$-axis gives result of spin-up, then measurement of $B$ along the $z$-axis will obtain result of spin-down. Similarly, if measurement of $A$ along the $y$-axis gives the result of spin-up, then measurement of $B$ along the $y$-axis will obtain the result of spin-down. Both measurements are mutually exclusive and are needed to provide a complete description of the entangled pair. The expectation value of Alice measuring $A$ along direction of unit vector $a$ and obtaining spin-up, and Bob measuring $B$ along direction of unit vector $b$ and obtaining spin-up, is denoted as $E(+,+|a, b, \Tilde{p})$ where $\Tilde{p}$ is the initial joint probability density for a singlet state. Since $\Tilde{p}$ comprises two mutual exclusive components given by \eqref{measABz} and \eqref{measABy}, we have
\begin{equation}
    \label{mutualExcl}
    E(+,+|a, b, \Tilde{p}) = E(+,+|a, b, \Tilde{p}_z) + E(+,+|a, b, \Tilde{p}_y)
\end{equation}

First we consider the initial condition described by the joint probability density functions given by \eqref{measABz}. This initial condition can be further decomposed such that half of the chance that the joint probability density $\Tilde{p}_{z,1}(\theta_A^z, \theta_B^z) = \delta(\theta_A^z)\delta(\theta_B^z-\pi)$ and the other half chance that $\Tilde{p}_{z,2}(\theta_A^z, \theta_B^z) = \delta(\theta_A^z-\pi)\delta(\theta_B^z)$. Now for the initial condition $\Tilde{p}_{z,1}(\theta_A^z, \theta_B^z) = \delta(\theta_A^z)\delta(\theta_B^z-\pi)$, suppose that Alice measures $A$ along a direction determined by unit vector $a$ which forms an angle with the $z$-axis by $\alpha$, and Bob measures $B$ along a direction determined by unit vector $b$ which forms an angle with $z$-axis by $\beta$. In typical Bell experiments, the measurements of Alice and Bob are performed almost at the same time. Therefore, we will not consider the time dependency of measurement outcomes as discussed in Section \ref{sec:SpinEnt}. 

The probability density of measurement outcome for Alice is given by \eqref{pinftyprime3} as
\begin{equation}
\label{Alice:a1}
    \bar{p}(\theta^z_a|\alpha) =\cos^2(\alpha/2)\delta(\theta^z_a) + \sin^2(\alpha/2)\delta(\theta^z_a-\pi),
\end{equation}
while the probability density of measurement outcome for Bob is given by
\begin{equation}
\label{Bob:b1}
    \bar{p}(\theta^z_b|\beta) =\sin^2(\beta/2)\delta(\theta^z_b) + \cos^2(\beta/2)\delta(\theta^z_b-\pi),
\end{equation}
Then, the statistical average of $S_A(a)S_B(b)$ for $(+,+)$, $(+,-)$, $(-,+)$, $(-,-)$ are
\begin{equation}
    \label{stat_ab_z}
    \begin{split}
        E(+,+|a,b,\Tilde{p}_z) &= \cos^2(\alpha/2)\sin^2(\beta/2)\\
        E(+,-|a,b,\Tilde{p}_z) &= -\cos^2(\alpha/2)\cos^2(\beta/2)\\
        E(-,+|a,b,\Tilde{p}_z) &= -\sin^2(\alpha/2)\sin^2(\beta/2)\\
        E(-,-|a,b,\Tilde{p}_z) &= \sin^2(\alpha/2)\cos^2(\beta/2).
    \end{split}
\end{equation}
The overall expectation value for this initial condition is
\begin{equation}
    \label{stat_ab_z}
    E(a, b|\Tilde{p}_z) = E(+,+|a,b,\Tilde{p}_z) + E(+,-|a,b,\Tilde{p}_z) + E(-,+|a,b,\Tilde{p}_z) + E(-,-|a,b,\Tilde{p}_z) = -\cos(\alpha)\cos(\beta).
\end{equation}
The initial condition can also be described by the joint probability density $\Tilde{p}_{z,2}(\theta_A^z, \theta_B^z) = \delta(\theta_A^z-\pi)\delta(\theta_B^z)$. Performing the same calculation with this initial condition, one will find that the resulting $E(a, b|\Tilde{p}_z)$ is the same. Thus, the statistical average of $E(a, b|\Tilde{p}_z)$ for both initial conditions is still given by \eqref{stat_ab_z}.

Next, we consider the initial condition described by the joint probability density functions given by \eqref{measABy}. This initial condition can be further decomposed such that half of the chance that the joint probability density $\Tilde{p}_{y,1}(\theta_A^y, \theta_B^y) = \delta(\theta_A^y)\delta(\theta_B^y-\pi)$ and the other half chance that $\Tilde{p}_{y,2}(\theta_A^y, \theta_B^y) = \delta(\theta_A^y-\pi)\delta(\theta_B^y)$. Now for the initial condition that $\Tilde{p}_{y,1}(\theta_A^y, \theta_B^y) = \delta(\theta_A^y)\delta(\theta_B^y-\pi)$, Alice's measurement direction of $a$ forms an angle with the $y$-axis given by $\alpha'=\pi/2 - \alpha$. Similarly, Bob's measurement direction of $b$ forms an angle with the $y$-axis given by $\beta'=\pi/2 - \beta$. The calculation for $E(a, b|y)$ is exactly the same as the calculation for $E(a, b|z)$ but with the replacement of $\alpha\to\alpha'$ and $\beta\to\beta'$. Thus,
\begin{equation}
    \label{stat_ab_y}
    E(a, b|\Tilde{p}_y) = -\cos(\alpha')\cos(\beta') = -\sin(\alpha)\sin(\beta).
\end{equation}
Similarly, the initial condition can also be described by the joint probability density $\Tilde{p}_{y,2}(\theta_A^y, \theta_B^y) = \delta(\theta_A^y-\pi)\delta(\theta_B^y)$, but the calculation result for $E(a, b|\Tilde{p}_y)$ is the same. Thus, the statistical average of $E(a, b|\Tilde{p}_y)$ for both initial conditions is still given by \eqref{stat_ab_y}.

According to \eqref{mutualExcl}, the overall expectation value $E(a, b)$ is obtained by adding \eqref{stat_ab_z} and \eqref{stat_ab_y},
\begin{equation}
    \label{stat_ab}
    E(a, b|\Tilde{p}) = E(a, b|\Tilde{p}_z) + E(a, b|\Tilde{p}_y) = -\cos(\alpha)\cos(\beta) -\sin(\alpha)\sin(\beta) = -\cos(\alpha - \beta).
\end{equation}
Let $\gamma = \alpha - \beta$, which is the angle formed by unit vectors $a$ and $b$. Thus, $E(a, b|\Tilde{p})=-\cos(\gamma) = -(a\cdot b)$ as desired.

Now consider the situation in which Bob delays his spin measurement on electron $B$ by a time period $\Delta t$. When $\Delta t$ is sufficiently large, even though Alice performs the measurement on $A$ and gets spin-up, Bob's measurement after $\Delta t$ can obtain spin-down or spin-up because the initial orientation for $B$ can be $\theta_B^z\in\Theta^+_z$ or $\theta_B^z\in\Theta^-_z$. Thus, Bob's measurement outcome can be 
\begin{equation}
\label{Bob:b2}
    \bar{p}(\theta^z_b|\beta) =\sin^2(\beta/2)\delta(\theta^z_b) + \cos^2(\beta/2)\delta(\theta^z_b-\pi),
\end{equation}
or, with equal probability,
\begin{equation}
\label{Bob:b3}
    \bar{p}(\theta^z_b|\beta) =\cos^2(\beta/2)\delta(\theta^z_b) + \sin^2(\beta/2)\delta(\theta^z_b-\pi).
\end{equation}
Therefore, the expectation value $E(a, b|\Tilde{p}_z)$ can be $E(a, b|\Tilde{p}_z)=-\cos(\alpha)\cos(\beta)$, or $E(a, b|\Tilde{p}_z)=\cos(\alpha)\cos(\beta)$ with the same probability. This results in the average $E(a, b|\Tilde{p}_z)=0$. Suppose that the correlation between $\theta_A^y$ and $\theta^y_B$ still holds, so that \eqref{stat_ab_y} is still valid. The overall expectation value becomes
\begin{equation}
    \label{stat_ab2}
    E(a, b|\Tilde{p}) = E(a, b|\Tilde{p}_z) + E(a, b|\Tilde{p}_y) = -\sin(\alpha)\sin(\beta).
\end{equation}
For the typical chosen experiment configurations $a=0$ and $a'=\pi/2$, and Bob chooses $b=\pi/4$ and $b'=3\pi/4$, one can calculate that 
\begin{equation}
    \label{CHSHDelay}
    | E(a, b) - E(a, b') + E(a', b) + E(a', b')| =\sqrt{2} < 2.
\end{equation}
Thus, the Bell-CHSH inequality is not violated.

If the correlation between $\theta_A^y$ and $\theta^y_B$ does not hold either after delay $\Delta t$, and $\theta_B^y\in\Theta^+_y$ or $\theta_B^y\in\Theta^-_y$ randomly, then one can calculate $E(a, b|\Tilde{p}_y)=0$. This results in $E(a, b|\Tilde{p}) = 0$ regardless of the configuration of $a$ and $b$. Clearly, in such a scenario, the Bell-CHSH inequality is not violated.

\end{document}